\documentclass[aps,UTF8,pra,floatfix,superscriptaddress]{revtex4}

\usepackage{mathrsfs}
\usepackage{graphicx}
\usepackage[english]{babel}
\usepackage{amsmath}
\usepackage{amssymb}
\usepackage{xcolor}
\usepackage{cancel}
\usepackage[normalem]{ulem}

\usepackage{relsize}
\usepackage{CJK}
\usepackage{dsfont}
\usepackage{bm}

\newcommand{\me}{\mathrm{e}}
\newcommand{\mi}{\mathrm{i}}

\newcommand{\dif}{\mathrm{d}}

\allowdisplaybreaks

\begin{document}

\title{Mathematical Foundation of the U$^N(1)$ Quantum Geometric Tensor}

\author{Xin Wang}
\affiliation{School of Physics, Southeast University, Jiulonghu Campus, Nanjing 211189, China}

\author{Xu-Yang Hou}
\affiliation{School of Physics, Southeast University, Jiulonghu Campus, Nanjing 211189, China}

\author{Jia-Chen Tang}
\affiliation{School of Physics, Southeast University, Jiulonghu Campus, Nanjing 211189, China}

\author{Hao Guo}
\email{guohao.ph@seu.edu.cn}
\affiliation{School of Physics, Southeast University, Jiulonghu Campus, Nanjing 211189, China}
\affiliation{Hefei National Laboratory, Hefei 230088, China}

\begin{abstract}
In this paper, we systematically establish the mathematical foundation for the $\text{U}^N(1)$ quantum geometric tensor (QGT) of mixed states
Explicitly, we present a description based on the $\text{U}^N(1)$ principal bundle and derive a Pythagorean-like distance decomposition equation.
Additionally, we offer a comprehensive comparison of its properties with those of the U(1) principal bundle description of the pure-state QGT. Finally, we prove a fundamental inequality for the $\text{U}^N(1)$ QGT and discuss its physical implication.
\end{abstract}

\maketitle

\section{Introduction}

Over the past few decades, the quantum geometric tensor (QGT)  \cite{QGTCMP80,QGT10} has emerged as a pivotal concept in the study of quantum states, providing valuable insights into their geometric and topological properties. The real part of QGT is the Fubini-Study metric, which quantifies the local distances in the state manifold, while the imaginary part relates to the Berry curvature, linking geometry with topology.
Research has shown that the QGT is capable of capturing the sensitivity of quantum systems to local perturbations, allowing for the differentiation of various quantum states through its complex structure. Its applications have expanded across multiple fields, including quantum statistical mechanics \cite{PhysRevLett.72.3439,KOLODRUBETZ20171}, quantum information \cite{IG_Book,doi:10.1142/S0219887824400115}, condensed matter physics \cite{PhysRevB.104.045103,PhysRevB.97.041108,PhysRevB.103.014516,PhysRevB.108.094508,PhysRevB.87.245103,PhysRevX.10.041041,Bhattacharya24}, PT-symmetric systems~\cite{PhysRevA.99.042104}, atomic, molecular, and optical (AMO) physics \cite{PhysRevB.94.134423,PhysRevB.91.214405,PhysRevA.97.033625}, as well as other disciplines \cite{cmp/1103904831,RevModPhys.82.1959,PhysRevResearch.3.L042018,PhysRevB.74.085308,PhysRevB.103.205415, PhysRevB.102.155407}. A variety of experimental techniques have been employed to explore the QGT, ranging from Rabi oscillation of an NV center in diamond~\cite{10.1093/nsr/nwz193}, quench or periodic driving of a superconducting qubit~\cite{PhysRevLett.122.210401}, photoluminescence studies of exciton-photon polaritons~\cite{Gianfrate2020}, Bloch state tomography of cold atoms~\cite{Yi23}, and transmission measurements of plasmonic lattices~\cite{Cuerda23}. Additionally, several other proposals have been put forward to further investigate this novel concept \cite{PhysRevB.97.201117,PhysRevLett.121.020401,PhysRevB.97.195422,PhysRevLett.124.197002}.

To date, most studies of the QGT have concentrated on pure states. However, generalizing it to mixed quantum states is both necessary and inevitable, as mixed states are far more prevalent in nature, especially in finite-temperature systems at thermal equilibrium. In developing the formalism of the QGT for pure states, it is necessary to eliminate extra gauge degrees of freedom, such as the phase factor, because two pure states that differ only by a phase factor are considered physically equivalent. Consequently, the pure-state QGT is invariant under local U(1) transformations. When extending this concept to mixed states, a similar requirement must be met to ensure that the corresponding QGT can properly measure the real distance between inequivalent mixed states. In our previous work (Ref.~\cite{PhysRevB.110.035144}), we developed a generalized QGT based on the Uhlmann connection of the mixed-state manifold.
With the help of the purification of density matrices, we extracted the U($N$) phase factors of mixed states via the polar decomposition. The mixed-state QGT was then derived by eliminating the extra gauge redundancy from the metric of the purification space. This  U($N$)-invariant QGT can effectively capture the  geometry of the mixed-state manifold through a local distance decomposition \cite{PhysRevB.110.035144}. Notably, its real part coincides with the Bures metric, which reduces to the Fubini-Study metric in the zero temperature limit. However, its imaginary part vanishes for ordinary physical processes, possibly due to the overly restrictive nature of the U($N$) gauge invariance.

In our second approach (Ref.~\cite{PhysRevB.110.035404}), we proposed an alternative QGT inspired by the Sj\"oqvist distance \cite{Andersson_2014,PhysRevResearch.2.013344}. In this framework, the requirement for gauge invariance is relaxed, which may enhance its experimental feasibility. The new QGT is invariant under the local $\underbrace{\text{U}(1)\times\cdots \times\text{U}(1)}_N=\text{U}^N(1)$ gauge transformation, and we refer to it as the $\text{U}^N(1)$ QGT. Its real part is a Riemannian metric, incorporating contributions from both the Fisher-Rao metric and the Fubini-Study metric. Meanwhile, its imaginary part introduces a nonzero U$^N(1)$-invariant 2-form. Additionally, it exhibits very interesting features in certain physical models \cite{PhysRevB.110.035404}. In this paper, we are going to systematically construct the mathematical foundation of the $\text{U}^N(1)$ QGT, presenting a description based on the $\text{U}^N(1)$ principal bundle and providing a Pythagorean-like equation for distance decomposition. Moreover, we find that the $\text{U}^N(1)$ QGT satisfies a fundamental inequality, similar to that of its pure-state counterpart.

The rest of the paper is organized as follows. In Sec.\ref{I}, we first briefly review the formalism of the pure-state QGT and then develop a systematic description based on the U(1) principal bundle, which facilitates a direct comparison for the later mathematical foundation of the mixed-state QGT. In Sec.\ref{II}, we first  present a detailed description of the mathematical foundation of the $\text{U}^N(1)$ QGT. Then, we prove a fundamental inequality, and discuss its physical implication. We also illustrate two examples exhibiting opposite properties of the $\text{U}^N(1)$ QGT. In the end, we conclude our findings in Sec.\ref{III}.

\section{The quantum distance and QGT for pure states}\label{I}

\subsection{Basic formalism}
The concept of physical distance between quantum states is both intriguing and significant in modern physics. Even when considering pure quantum states, this problem is more complex than one might initially expect. We examine a family of normalized states $|\psi(\mathbf{R})\rangle$, which depends on a set of continuous real parameters $\mathbf{R}=(R^1,R^2,\cdots, R^k)^T$, where $\mathbf{R}$ spans the parameter manifold $\mathcal{M}$.
A typical example of $\mathcal{M}$ is the Brillouin zone of a lattice system.
Here, $\mathbf{R}$ can be viewed as a local coordinate. Consequently, the (squared) local distance between quantum states with respect to variations in  $|\psi(\mathbf{R})\rangle$ is
\begin{align}\label{qd1}
\dif s^2=\big||\psi(\mathbf{R}+\dif \mathbf{R})\rangle-|\psi(\mathbf{R})\rangle\big|^2=\sum_{\mu\nu}\langle \partial_\mu\psi|\partial_\nu\psi\rangle\dif R^\mu\dif R^\nu.
\end{align}
However, this is not a proper definition of quantum distance since it is not invariant under the local U(1) transformation: $|\psi(\mathbf{R})\rangle\rightarrow \me^{\mi\chi(\mathbf{R})}|\psi(\mathbf{R})\rangle$. Hence, we refer to it as the `raw distance' between quantum states.
A straightforward evaluation reveals that the distance changes as follows:
\begin{align}\label{qd2}
\dif s^2\rightarrow \dif s'^2=\left(\langle\partial_\mu\psi|\partial_\nu\psi\rangle-\mi \omega_\mu\partial_\nu\chi-\mi \omega_\nu\partial_\mu\chi+\partial_\mu\chi\partial_\nu\chi\right)\dif R^\mu\dif R^\nu,
\end{align}
where \begin{align}\label{BA1}\omega_\mu=\langle\psi|\partial_\mu\psi\rangle=-\langle\partial_\mu\psi|\psi\rangle.\end{align}
 Interestingly, $\omega_\mu$ is the well-known Berry connection, which transforms like a gauge potential under this U(1) transformation: $\omega_\mu\rightarrow \omega_\mu'=\omega_\mu+\mi\partial_\mu\chi$. Therefore, we can redefine a `proper quantum distance' based on the properties of the gauge potential \cite{QGTCMP80}:
 \begin{align}\label{qd3}
\dif s^2
=\left(\langle\partial_\mu\psi|\partial_\nu\psi\rangle+\omega_\mu\omega_\nu\right)\dif R^\mu\dif R^\nu=\left(\langle\partial_\mu\psi|\partial_\nu\psi\rangle-\langle\psi|\partial_\mu\psi\rangle\langle\partial_\nu\psi|\psi\rangle\right)\dif R^\mu\dif R^\nu.
\end{align}
In this way, the change of $\omega_\mu$ compensates for the change in the original $\dif s^2$ in Eq.(\ref{qd2}), ensuring that the new distance is gauge invariant. The corresponding metric
 \begin{align}\label{qd4}
Q_{\mu\nu}=\langle\partial_\mu\psi|\partial_\nu\psi\rangle+\omega_\mu\omega_\nu=\langle\partial_\mu\psi|(1-|\psi\rangle\langle \psi|)|\partial_\nu\psi\rangle
\end{align}
is also referred to as the quantum geometric tensor (QGT) as it precisely captures the local geometric properties of quantum states.

\subsection{Fibre bundle description}\label{I.B}

 The previous discussion can be systematically framed within the context of a U(1) principal bundle, which provides a clear insight into the geometrical origin of the correction (\ref{qd3}). Suppose that all unnormalized quantum states under  consideration span an $N$-dimensional Hilbert space $\mathcal{H}$. The normalized states then form a unit sphere $S(\mathcal{H})$ in $\mathcal{H}$. Topologically, $S(\mathcal{H})$ is equivalent to $S^{2N-1}$, the unit sphere of real dimension $2N-1$. Different from the previous discussion, we use the tilde symbol $|\tilde{\psi}\rangle$ to label a normalized state. Two states are considered physically equivalent if they differ only by a U(1) phase factor, i.e. $|\tilde{\psi}_1\rangle\simeq|\tilde{\psi}_2\rangle$ if $|\tilde{\psi}_1\rangle=\me^{\mi\chi}|\tilde{\psi}_2\rangle$. The reason why the distance (\ref{qd1}) fails is that it still gives a non-zero result for two physically equivalent quantum states. By factoring out the phase factor, we obtain the equivalence classes of normalized states. Here, we use the untilded symbol $|\psi\rangle$ to denote a representative of the equivalence class $[|\tilde{\psi}\rangle]$, where $|\tilde{\psi}\rangle=\me^{\mi\theta}|\psi\rangle$ for some $\me^{\mi\theta}\in$U(1). If $|\psi_1\rangle\neq |\psi_2\rangle$, they represent two physically inequivalent states. Consequently, all such states form the quotient space $S^{2N-1}/\text{U}(1)=CP^{N-1}$, which is the well-known Hopf fibration with $CP^{N-1}$ being the $(N-1)$-dimensional complex projective space.
 We also refer to $CP^{N-1}$ as the quantum phase space since all extra degrees of freedom have been eliminated.
 Note that Eqs.(\ref{qd1}) and (\ref{qd3}) actually represent the distances on $S^{2N-1}$ and $CP^{N-1}$, respectively. Then, what is the relation between them? Here, we present the fibre bundle description that underlies this relationship.

 We take $S^{2N-1}$ as the total space, and $CP^{N-1}$ as the base manifold. A canonical projection connecting them is given by $\pi: S^{2N-1}\rightarrow CP^{N-1}$, which collapses all phase factors to project out the physically inequivalent classes of states: $\pi(|\tilde{\psi}\rangle)\equiv\pi(\me^{\mi\theta}|\psi\rangle)=|\psi\rangle$. Conversely,
  \begin{align}
  \pi^{-1}(|\psi\rangle)=\{\me^{\mi\theta}|\psi\rangle|\me^{\mi\theta}\in\text{U}(1)\}
\end{align}
represents the fibre space $F_{|\psi\rangle}$ at the point $|\psi\rangle\in CP^{N-1}$. For arbitrary $|\tilde{\psi}\rangle\in S^{2N-1}$, the right action $R_{|\tilde{\psi}\rangle}$ maps $g\in\text{U}(1)$ to:
$R_{|\tilde{\psi}\rangle}(g)=|\tilde{\psi}\rangle g=g|\tilde{\psi}\rangle\in \pi^{-1}(\pi(|\tilde{\psi}\rangle))=F_{|\psi\rangle}$.
This implies $R_{|\tilde{\psi}\rangle}[\text{U}(1)]=F_{|\psi\rangle}$. Since the right action is a one-to-one mapping, the fiber space is then isomorphic to the structure group U(1), indicating that this fibre bundle is a principal bundle.
A smooth map $\sigma: CP^{N-1}\rightarrow S^{2N-1}$ given by $\sigma|\psi(\mathbf{R})\rangle=\me^{\mi\theta(\mathbf{R})}|\psi(\mathbf{R})\rangle$ is called a section, which satisfies $\pi\circ \sigma=\text{id}$. Clearly, a section locally fixes the phase factor of a state $|\psi\rangle$.
Under the current terminology, the connection defined by Eq.(\ref{BA1}) is, in fact, a U(1) connection on the total space $S^{2N-1}$, given by $\omega=\langle\tilde{\psi}|\dif |\tilde{\psi}\rangle$. Substituting $|\tilde{\psi}\rangle=\me^{\mi\theta}|\psi\rangle$, we obtain
  \begin{align}\label{omega1}
 \omega=\mi\dif \theta+\mathcal{A}=\mi(\dif\theta-\mi\mathcal{A}),
\end{align}
where $\mi\dif \theta$ is a connection on the fibre space, and $\mathcal{A}=\langle \psi|\dif|\psi\rangle$ is the Berry connection on the base manifold $CP^{N-1}$. To understand the role of $\omega$, we consider a smooth loop $\mathbf{R}(t)$ in $\mathcal{M}$, where $0\le t\le \tau$ and $\mathbf{R}(0)=\mathbf{R}(\tau)$. It introduces a loop $\gamma(t):=|\psi(t)\rangle\equiv|\psi(\mathbf{R}(t))\rangle$ in $CP^{N-1}$. The section $\sigma$ defines a lift of $\gamma$ as $\tilde{\gamma}(t):=|\tilde{\psi}(t)\rangle=\sigma(|\psi(t)\rangle)=\me^{\mi\theta(t)}|\psi(t)\rangle\equiv\me^{\mi\theta(\mathbf{R}(t))}|\psi(\mathbf{R}(t))\rangle$, which satisfies $\pi\circ \tilde{\gamma}=\gamma$. Note that $\tilde{\gamma}(t)$ is not necessarily a closed curve, as generically $\theta(0)\neq \theta(\tau)$. $|\tilde{\psi}(t)\rangle$ is said to be parallel transported if it satisfies the condition
  \begin{align}\label{pc0}
\langle\tilde{\psi}(t)|\frac{\dif}{\dif t}  |\tilde{\psi}(t)\rangle=0\Rightarrow \mi\frac{\dif
\theta(t)}{\dif t}+\langle \psi(t)|\frac{\dif}{\dif t}|\psi(t)\rangle=0,
\end{align}
which ensures that $|\tilde{\psi}(t+\dif t)\rangle$ remains in phase with $|\tilde{\psi}(t)\rangle$. Solving this equation, we get the Berry phase at the end of the transportation: $\theta(\tau)=\mi\mathlarger{\int}_0^\tau\dif t\langle\psi(t)|\dot{\psi}(t)\rangle$, which is invariant under the U(1) structure group. In this context, we refer to $\tilde{\gamma}$ as the horizontal lift of $\gamma$. Let $\tilde{X}$ and $X$ be the tangent vectors to $\tilde{\gamma}$ and $\gamma$, respectively. The vector $X$ is the pushforward of $\tilde{X}$ by $\pi$: $X=\pi_*\tilde{X}$. Furthermore, the parallel-transport condition (\ref{pc0}) can be reexpressed as
  \begin{align}\label{pca1}\omega(\tilde{X})=0.
\end{align}
In other words, $\tilde{X}$ is the horizontal vector that belongs to the horizontal subspace of the tangent space $TS^{2N-1}$, and the one-form $\omega$ decomposes $TS^{2N-1}$ into its vertical and horizontal subspaces. Therefore, $\omega$ is the Ehresmann connection on $S^{2N-1}$. The Berry connection is the pullback of $\omega$ by $\sigma$: $\mathcal{A}=\sigma^*\omega$.

Building on this foundation, we can now grasp the geometric basis underlying the gauge correction to the distance discussed in the previous section.
According to Eq.(\ref{qd1}), the distance on the total space can be separated into
  \begin{align}\label{d2}
  \dif s^2(S^{2N-1})&=\langle \dif \tilde{\psi}|\dif \tilde{\psi}\rangle=
  \langle \dif \psi|\dif \psi\rangle-2\mi\dif \theta \mathcal{A}+(\dif \theta)^2\notag\\
  &=  \langle \dif \psi|\dif \psi\rangle+\mathcal{A}^2-\left(\mi\dif\theta+\mathcal{A}\right)^2\notag\\
  &=\dif s^2(CP^{N-1})+\left|\omega\right|^2,
\end{align}
where
  \begin{align}\label{d3}
 \dif s^2(CP^{N-1})=\langle \dif \psi|\dif \psi\rangle+\mathcal{A}^2=\sum_{\mu\nu}\left(\langle\partial_\mu\psi|\partial_\nu\psi\rangle+\mathcal{A}_\mu\mathcal{A}_\nu\right)\dif R^\mu\dif R^\nu=
 \sum_{\mu\nu}\langle\partial_\mu\psi|(1-|\psi\rangle\langle \psi|)|\partial_\nu\psi\rangle\dif R^\mu\dif R^\nu
\end{align}
denotes the local distance on $CP^{N-1}$, and $|\omega|^2=-\left(\mi\dif\theta+\mathcal{A}\right)^2$ represents the local distance on the fibre space. Note that the absolute-value sign arises since $\omega$, $\mi\dif\theta$ and $\mathcal{A}$ are all purely imaginary. Eq.(\ref{d2}) provides a Pythagorean-like decomposition of the local distance on the total space.
If a quantum state is parallel transported along a smooth curve, then $\dif s^2(S^{2N-1})=\dif s^2(CP^{N-1})$, indicating that the local distance on the fiber space does not contribute to the total distance. Equivalently, the minimization condition of $\dif s^2(S^{2N-1})$ is the parallel-transport condition (\ref{pca1}).
This is reasonable since no instantaneous phase is generated during a parallel transport. From Eq.(\ref{d3}), it follows that the metric tensor on the base manifold thus reproduces the gauge-invariant QGT in Eq.(\ref{qd4}):
\begin{align}\label{qd5}
Q_{\mu\nu}=\langle\partial_\mu\psi|\partial_\nu\psi\rangle+\mathcal{A}_\mu\mathcal{A}_\nu=\langle\partial_\mu\psi|(1-|\psi\rangle\langle \psi|)|\partial_\nu\psi\rangle.
\end{align}
Its real part is the well-known Fubini-Study metric:
\begin{align}
g^\text{FS}_{\mu\nu}=\frac{1}{2}\left(\langle\partial_\mu\psi|\partial_\nu\psi\rangle+\langle\partial_\nu\psi|\partial_\mu\psi\rangle\right)-\langle\partial_\mu\psi|\psi\rangle\langle \psi|\partial_\nu\psi\rangle,\end{align}
 and its (negative) imaginary part \begin{align}\Omega_{\mu\nu}=\frac{\mi}{2}\left(\langle\partial_\mu\psi|\partial_\nu\psi\rangle-\langle\partial_\nu\psi|\partial_\mu\psi\rangle\right)=\frac{\mi}{2}\mathcal{F}_{\mu\nu}\end{align} is proportional to the Berry curvature.


\section{The Sj\"oqvist distance and the U$^N(1)$ quantum geometric tensor}\label{II}

\subsection{Overview of the Basic formalism}

The U$^N(1)$ quantum geometric tensor (QGT), also known as the Sj$\ddot{\text{o}}$qvist QGT, is derived from the Sj\"oqvist quantum distance between full-ranked density matrices. Similarly, we consider a smooth path $\mathbf{R}(t)=(R^1(t),R^2(t),\cdots, R^k(t))^T$ in $\mathcal{M}$. This introduces an evolving mixed state $\rho(t)\equiv \rho(\mathbf{R}(t))$. Let $|n(t)\rangle$ ($n=0,1,\cdots$ $N-1$) be
 the instantaneous eigenstate of $\rho(t)$. Then, $\rho(t)$ can be diagonalized as $\rho(t)=\sum_{n=0}^{N-1}\lambda_n(t)|n(t)\rangle\langle n(t)|$.
Following Ref.~\cite{PhysRevResearch.2.013344}, we further introduce $N$ spectral rays $\{\me^{\mi\theta_n(t)}|n(t)\rangle\}$ ($n=0,1,\cdots$ $N-1$) along the path $\mathbf{R}(t)$ and
let $\mathcal{B}(t)=\{\sqrt{\lambda_n(t)}\{\me^{\mi\theta_n(t)}|n(t)\rangle\}_{n=0}^{N-1}$ be the spectral decomposition along the path.
The Sj$\ddot{\text{o}}$qvist distance is defined as the minimum distance between $\mathcal{B}(t)$ and $\mathcal{B}(t+\dif t)$:
\begin{widetext}
\begin{align}\label{Sdis}
\dif^2_\text{S}(t+\dif t,t)=&\inf_{\theta_n}\sum_{n=0}^{N-1}\big|\sqrt{\lambda_n(t+\dif t)}\me^{\mi\theta_n(t+\dif t)}|n(t+\dif t)\rangle-\sqrt{\lambda_n(t)}\me^{\mi\theta_n(t)}|n(t)\rangle\big|^2 \notag \\
=&2-2\sup\sum_n\sqrt{\lambda_n(t)\lambda_n(t+\dif t)}|\langle n(t)|n(t+\dif t)\rangle|\left\{\dot{\theta}_n(t)\dif t+\arg\left[1+\langle n(t)|\dot{n}(t)\rangle \dif t\right]+O(\dif t^2)\right\}.
\end{align}
\end{widetext}
The infimum is taken among all possible sets of spectral phases $\{\theta_n(t),\theta_n(t+\dif t)\}$, subject to the condition $\dot{\theta}_n(t)\dif t+\arg \langle n(t)|n(t+\dif t)\rangle=0$. Since $\arg \langle n(t)|n(t+\dif t)\rangle\approx-\mi\langle n(t)|\dot{n}(t)\rangle\dif t$,
the minimization condition is equivalent to
\begin{align}\label{Sdispcb}
\mi\dot{\theta}_n(t)+\langle n(t)|\dot{n}(t)\rangle=0, \quad \text{for} \quad n=0,\cdots, N-1,
\end{align}
which corresponds precisely to the parallel transport condition for each individual pure state in the ensemble. Under this condition, the Sj\"oqvist distance becomes
\begin{align}\label{Sjodis2}
\dif^2_\text{S}(t ,t+\dif t)
=&2-2\sum_n\sqrt{\lambda_n}\left[\sqrt{\lambda_n}+\frac{\dot{\lambda_n}}{2\sqrt{\lambda_n}}\dif t -\frac{1}{2}\left(\frac{\dot{\lambda_n}^2}{4\lambda_n^{\frac{3}{2}}}-\frac{\ddot{\lambda}_n}{2\sqrt{\lambda_n}}\right)\dif t^2+O(\dif t^3)\right]\notag\\
\times& \sqrt{1+(\langle n|\dot{n}\rangle+\langle \dot{n}|n\rangle)\dif t+\left[\langle \dot{n}|n\rangle\langle n|\dot{n}\rangle+\frac{1}{2}(\langle n|\ddot{n}\rangle+\langle \ddot{n}|n\rangle)\right]\dif t^2+O(\dif t^3)}\notag\\
=&\sum_n\left[\frac{\dot{\lambda_n}^2}{4\lambda_n}+\lambda_n\langle\dot{n}|(1-|n\rangle\langle n|)|\dot{n}\rangle\right] \dif t^2,
\end{align}
where the following conditions have been applied: $\sum_n\dot{\lambda}_n=\sum_n\ddot{\lambda}_n=0$, $\langle n|\dot{n}\rangle+\langle \dot{n}|n\rangle=0$, and $\langle n|\ddot{n}\rangle+\langle \ddot{n}|n\rangle=-2\langle \dot{n}|\dot{n}\rangle$. In terms of the parameters $\mathbf{R}$, the Sj\"oqvist distance can also be expressed as $\dif^2_\text{S}(\mathbf{R} ,\mathbf{R}+\dif \mathbf{R})=Q^\text{S}_{\mu\nu}\dif R^\mu\dif R^\nu$ with
\begin{align}\label{Sm0}
Q^\text{S}_{\mu\nu}=\sum_n\left[\frac{\partial_\mu\lambda_n\partial_\nu\lambda_n}{4\lambda_n}+\lambda_n\langle\partial_\mu n|(1-|n\rangle\langle n|)|\partial_\nu n\rangle\right].
\end{align}
The first term represents the Fisher-Rao metric, while the second term is the weighted sum of the QGT for each spectral ray, as given in Eq.(\ref{qd5}).
Note that the Sj\"oqvist distance is calculated by taking the minimum across all possible sets of spectral phases, ensuring that it remains invariant under local gauge transformations of the form $\mathcal{U}(\mathbf{R})=\text{diag}(\me^{\mi\chi_0(\mathbf{R})},\cdots,\me^{\mi\chi_{N-1}(\mathbf{R})})\in \text{U}^N(1)$. Indeed, this distance measures the real quantum distance between `adjacent' mixed states in terms of spectral rays. Consequently, $Q^\text{S}_{\mu\nu}$ also remains invariant under this local U$^N(1)$ transformation. Since it captures the local properties of quantum states, it is also referred to as the U$^N(1)$ QGT.

\subsection{Mathematical foundation for U$^N(1)$ QGT }

\subsubsection{Purification of density matrix}

Similarly, the formalism of the U$^N(1)$ QGT can also be systematically derived using the language of fiber bundles. The key point is to  identify the counterpart of a pure-state wavefunction in the context of mixed states. This can be accomplished by decomposing a $N$-dimensional density matrix as $\rho=WW^\dag$, where $W$ is referred to as the purification of $\rho$ and serves a role analogous to that of the wavefunction.  Conversely, the polar decomposition of $W$ is $W=\sqrt{\rho}U$, where $U\in$U$(N)$ represents the phase factor of $W$.  It is important to note that this decomposition is unique only when $\rho$ is full-rank.  Clearly, $U$ generalizes the U(1) phase factor $\me^{\mi\chi}$ from pure states to mixed states. Furthermore, $W$ also has a pure-state-like representation, the purified state $|W\rangle$. If $\rho$ is diagonalized as $\rho=\sum_n\lambda_n|n\rangle\langle n|$, then $W=\sum_n\sqrt{\lambda_n}|n\rangle\langle n|U$, and the purified state is $|W\rangle=\sum_n\sqrt{\lambda_n}|n\rangle\otimes U^T|n\rangle$. In other words, $|W\rangle$ is obtained by formally applying the transpose operation to $\langle n|U$ in $W$. Finally, the Hilbert-Schmidt inner product between two purifications/purified states is defined as: $\langle W_1,W_2\rangle=\langle W_1|W_2\rangle=\text{Tr}(W^\dag_1W_2)$, which introduces a norm $||W||^2=\langle W,W\rangle$. Since $\text{Tr}\rho=1$, then $||W||=\sqrt{\text{Tr} (W^\dag W)}=\sqrt{\text{Tr} (WW^\dag)}=1$.

 It is known that the density matrix is not in one-to-one correspondence with mixed states, as stated by Schr\"odinger's mixture theorem\cite{Bengtsson_book}. However, the density matrix is sufficient to predict any physical measurement in its associated mixed state. For an observable $\mathcal{O}$, its expectation value in this mixed state is determined by $\langle \mathcal{O}\rangle=\text{Tr}(\rho \mathcal{O})$. Thus, two different density matrices, $\rho_1$ and $\rho_2$, correspond to two physically inequivalent mixed states since there must exist at least one observable $\tilde{\mathcal{O}}$ such that $\text{Tr}(\rho_1 \tilde{\mathcal{O}})\neq \text{Tr}(\rho_2 \tilde{\mathcal{O}})$. A density matrix can have infinitely many purifications, all of which correspond to a single physically equivalent mixed state. Let $\rho$ and its purification $W$ continuously depend on the parameter $\mathbf{R}$. The distance between nearby purifications with respect to variations of $\mathbf{R}$ is given by
 \begin{align}\label{ds2}
\dif s^2=\big|| W(\mathbf{R}+\dif \mathbf{R})\rangle-|W(\mathbf{R})\rangle\big|^2=\langle \partial_\mu W|\partial_\nu W\rangle\dif R^\mu\dif R^\nu=\text{Tr}(\partial_\mu W^\dag\partial_\nu W)\dif R^\mu\dif R^\nu.
\end{align}
Similar to Eq.(\ref{qd1}), this is the `raw distance' between mixed states. For comparison, the Sj\"oqvist distance in the previous section is invariant under local $\text{U}^N(1)$ gauge transformations, which allows it to be recognized as a proper distance between certain types of inequivalent mixed states.
To obtain the Sj\"oqvist distance by eliminating extra gauge redundancies from the raw distance, a systematic description based on fiber bundles is required, analogous to the approach for pure states.

\subsubsection{Fibre bundle description}

To construct a fibre bundle description, we first denote $\mathcal{D}_N^N$ as the space spanned by all full-rank density matrices of dimension $N$, i.e., $\forall \rho\in\mathcal{D}^N_N$, rank$(\rho)=N$. Mathematically, $\mathcal{D}_N^N$ is a manifold equipped with a Riemannian metric\cite{Bengtsson_book}, and it can also be referred to as the phase space of mixed states \cite{PhysRevB.110.035144}. To generate the Sj\"oqvist distance, we select a special purification of $\rho=\sum_n\lambda_n|n\rangle\langle n|$, given by
\begin{align}\label{W}
W=\sum_n\sqrt{\lambda_n}|n\rangle\langle n_0|\me^{\mi\theta_n}.\end{align} Here, $\{|n_0\rangle\}_{n=0}^{N-1}$ is a set of fixed states, and their specific selection will be detailed in the following discussion.
Comparing with $W=\sqrt{\rho}U$, the phase factor is given by
\begin{align}
U =\sum_n\me^{\mi\theta_n }|n \rangle\langle n_0|\sim
\begin{pmatrix}
\me^{\mi\theta_0 } & &\\ & \ddots & \\ & & \me^{\mi\theta_{N-1} }
\end{pmatrix}.
\end{align}
Thus, the set of all phase factors of a single $W$ is isomorphic to U$^N(1)$.
All purifications given in Eq.(\ref{W}) form the total space $S_N=\{W|W=\sqrt{\rho}U, \rho\in\mathcal{D}^N_N, U\in \text{U}^N(1),\text{ and }||W||=1\}$, and the fibre space at a point $\rho$ is $F_\rho\sim \text{U}^N(1)$.
The corresponding fibration is $S_N/\text{U}^N(1)=\mathcal{D}^N_N$.
Similar to the previous approach, we assume that $\rho$ continuously depends on the parameter $\mathbf{R}$, i.e. $\rho(\mathbf{R})=\sum_n\lambda_n(\mathbf{R})|n(\mathbf{R})\rangle\langle n(\mathbf{R})|$.
In this context, $\mathbf{R}$ serves as the local coordinate of $\mathcal{D}^N_N$. We select a fixed point $\mathbf{R}_0$ and define $|n_0\rangle=|n(\mathbf{R}_0)\rangle$.
Next, we introduce the local gauge transformation $\mathcal{U}(\mathbf{R})=\sum_n|n_0\rangle\langle n_0|\me^{\mi\chi_n(\mathbf{R})}$, which also form a U$^N(1)$ group. Let $W\in F_\rho$. Under the right action $W'=W\mathcal{U}$, the gauge transformed purification $W'$  remains in the the same fibre space $F_\rho$. This indicates that all $\mathcal{U}$ span the structure group U$^N(1)$, which is isomorphic to $F_\rho$. Consequently, we can construct a U$^N(1)$ principal bundle.
The canonical projection is defined as $\pi: S_N\rightarrow \mathcal{D}^N_N$ such that $\pi(W)=WW^\dag=\rho$. Conversely, the smooth map
$\sigma(\rho(\mathbf{R}))=\sum_n\sqrt{\lambda_n(\mathbf{R})}|n(\mathbf{R})\rangle\langle n_0|\me^{\mi\theta_n(\mathbf{R})}$ defines a section. Note that there always exists a global section $\sigma(\rho)=\sqrt{\rho}$, which means the phase factor is trivial. Accordingly, this principal bundle is topologically trivial, distinguishing it from the case of pure states.

To match the minimization condition (\ref{Sdispcb}), we introduce an Ehresmann connection on $S_N$:
\begin{align}
\omega=\text{Tr}\left(W^\dag \dif W\right)=\text{Tr}\left[\sum_n\left(\sqrt{\lambda_n}\dif\sqrt{\lambda_n}+\lambda_n\mi\dif \theta_n \right)|n_0\rangle\langle n_0|+\sqrt{\lambda_m\lambda_n}\sum_{mn}\me^{\mi(\theta_m-\theta_n)}\langle n|\dif|m\rangle|n_0\rangle\langle m_0|\right],
\end{align}
Here the dependence on $\mathbf{R}$ is compressed for convenience.
Using the fact that $\sum_n\sqrt{\lambda_n}\dif\sqrt{\lambda_n}=\dif \sum_n \lambda_n=0$, we get
\begin{align}
\omega=\sum_n\lambda_n\left(\mi\dif \theta_n+\mathcal{A}_n\right)
\end{align}
where $\mathcal{A}_n=\langle n|\dif n\rangle$ is the Berry connection for the $n$-the level $|n\rangle$.
Similarly, the first and second terms on the right-hand-side represent the associated connections on the fibre space and the base manifold respectively. This formulation is a direct generalization of the pure-state case.
Under the local gauge transformation $W'=W\mathcal{U}$, $\omega$ transforms as $\omega'=\omega+\mi\sum_n\lambda_n\dif \chi_n$, thereby it is a U(1) connection. Consider a loop $\mathbf{R}(t)$ ($0\le t\le\tau$) in $\mathcal{M}$, which satisfies   $\mathbf{R}(0)=\mathbf{R}(\tau)=\mathbf{R}_0$. It also induces a loop $\gamma(t)=\rho(t)\equiv \rho(\mathbf{R}(t))$ in $\mathcal{D}^N_N$. The lift of $\gamma(t)$ is a curve in $S_N$: $\tilde{\gamma}(t)=\sigma(\rho(t))=W(t)\equiv W(\mathbf{R}(t))$, which is not necessarily closed. Denote the tangent vectors of $\gamma$ and $\tilde{\gamma}$ by $X$ and $\tilde{X}$, respectively. If $\tilde{\gamma}$ is a horizontal lift of $\gamma$, then $\omega(\tilde{X})=0$, which implies
\begin{align}\label{pxe}
\text{Tr}(W^\dag \dot{W})=\sum_n\lambda_n\left(\mi\dot{\theta}_n(t)+\mathcal{A}_n(\tilde{X})\right)=\sum_n\lambda_n\left(\mi\dot{\theta}_n(t)+\langle n(t)|\frac{\dif}{\dif t}|n(t)\rangle\right)=0.
\end{align}
If this condition is met, $W(t)$ is said to be parallel transported along $\mathbf{R}(t)$.
Comparing with Eq.(\ref{Sdispcb}), it can be found that the minimization condition for Sj\"oqvist distance is  a sufficient condition for this parallel-transport condition. Furthermore, the pullback of $\omega$ by $\sigma$ defines a U(1) connection on the base manifold:
\begin{align}
\mathcal{A}_{\mathcal{D}^N_N}=\sigma^*\omega=\sum_n\lambda_n\mathcal{A}_n.
\end{align}

In context of the U$^N(1)$ principal bundle, Eq.(\ref{ds2}) in fact represents the local distance on the total space $S_N$, i.e. the `raw distance' $\dif s^2 (S_N)$ between adjacent mixed states.
A straightforward evaluation shows that
\begin{align}\label{Sdis3c}
\dif s^2(S_N)=&\sum_n\left[\partial_\mu \sqrt{\lambda_n}\partial_\nu \sqrt{\lambda_n}+\lambda_n(\langle\partial_\mu n|\partial_\nu n\rangle+\partial_\mu \theta_n\partial_\nu\theta_n -\mi\mathcal{A}_{n\mu}\partial_\nu\theta_n-\mi\mathcal{A}_{n\nu}\partial_\mu\theta_n) \right]\dif R^\mu \dif R^\nu\notag\\
=&\sum_n\big\{(\dif \sqrt{\lambda_n})^2+\lambda_n\big[\langle\dif n|\dif n\rangle+(\dif\theta_n-\mi\mathcal{A}_n)^2+(\mathcal{A}_n)^2\big]\big\},
\end{align}
where $\mathcal{A}_{n\mu}=\langle n|\partial_\mu n\rangle=-\langle \partial_\mu n| n\rangle$ is the $n$-th component of $\mathcal{A}_n$, and the second line is expressed in terms of differential forms. Obviously, the raw distance is minimized when
 \begin{align}\label{pxc4}
 \partial_\mu\theta_n-\mi\mathcal{A}_{n\mu}=0,\quad \text{ or } \quad \dif\theta_n-\mi\mathcal{A}_n=0,
\end{align}
which exactly agrees with Eq.(\ref{Sdispcb}), the minimization condition of the Sj\"oqvist distance. Similar to Eq.(\ref{d2}), the raw distance can also be decomposed as
\begin{align}\label{Sdis3e}
\dif s^2(S_N)=\dif s^2 (\mathcal{D}^N_N)+\sum_n\lambda_n(\dif\theta_n-\mi\mathcal{A}_n)^2.
\end{align}
Here
\begin{align}\label{Sdis3f}
\dif s^2 (\mathcal{D}^N_N)=\sum_n\left\{(\dif \sqrt{\lambda_n})^2+\lambda_n\big[\langle\dif n|\dif n\rangle+(\mathcal{A}_n)^2\big]\right\}
=\sum_n\left[\frac{(\dif \lambda_n)^2}{4\lambda_n}+\lambda_n\langle \dif n|(1-|n\rangle\langle n|)|\dif n\rangle\right]
\end{align}
is the distance on the base manifold $\mathcal{D}^N_N$, which is also the Sj\"oqvist distance by comparing with Eq.(\ref{Sjodis2}). The second term $\sum_n\lambda_n(\dif\theta_n-\mi\mathcal{A}_n)^2$ represents the local distance on the fibre space.
This decomposition is entirely consistent with the fibration $S_N/\text{U}^N(1)=\mathcal{D}^N_N$. When the total distance $\dif s^2(S_N)$ is minimized to $\dif s^2 (\mathcal{D}^N_N)$, the associated purification must undergoes a parallel transport. However, this does not hold conversely, as the parallel transport condition is merely a necessary  implication of the minimization condition (\ref{Sdispcb}). This situation is slightly different from that of pure states.
As previously mentioned, $\dif s^2 (\mathcal{D}^N_N)$ or the Sj\"oqvist distance can be further separated into $\dif s^2 (\mathcal{D}^N_N)=\dif s^2_\text{FR}+\sum_n\lambda_n\dif s^2_{\text{FS}n}$, where $\dif s^2_\text{FR}=\sum_n\frac{(\dif\lambda_n)^2}{4\lambda_n}$ is the Fisher-Rao distance, and $\dif s^2_{\text{FS}n}=\langle \dif n|(1-|n\rangle\langle n|)|\dif n\rangle$ is the Fubini-Study distance for the $n$-th level.

\begin{table}[t]
\centering
\caption{A comparison between the geometries of pure and mixed states.
}\label{table1}
\begin{tabular}{|c|c|c|}\hline
  & Pure state & Mixed state \\ \hline
Total space 
 & $S^{2N-1}$ & $S_N$  \\ \hline
  Phase space & $CP^{N-1}$ & $\mathcal{D}^N_N$ \\\hline
  Fibration & $S^{2N-1}/\text{U}(1)=CP^{N-1}$ & $S_N/\text{U}^N(1)=\mathcal{D}^N_N$ \\\hline
  Connection & Berry connection $\mathcal{A}$ & Weighted sum of Berry connection $\sum_n\lambda_n\mathcal{A}_n$ \\\hline
  Raw distance & $\dif s^2(S^{2N-1})$ & $\dif s^2(S_N)$ \\\hline
  Gauge-invariant distance & $\dif s^2(CP^{N-1})$ & $\dif s^2(\mathcal{D}^N_N)$ \\\hline
  Relations between distances & $\dif s^2(S^{2N-1})=\dif s^2(CP^{N-1})+(\dif \theta-\mi\mathcal{A})^2$ & $\dif s^2(S_N)=\dif s^2 (\mathcal{D}^N_N)+\sum_n\lambda_n(\dif\theta_n-\mi\mathcal{A}_n)^2$ \\\hline
  Raw metric & $\langle \partial_\mu\tilde{\psi}|\partial_\nu\tilde{\psi}\rangle$ & $\langle \partial_\mu W|\partial_\nu W\rangle$\\\hline
  Real part of QGT & \text{Re}$\langle \partial_\mu\psi|\partial_\nu\psi\rangle+\mathcal{A}_\mu\mathcal{A}_\nu$ (Fubini-Study) & $\sum_n\left[\frac{\partial_\mu\lambda_n\partial_\nu\lambda_n}{4\lambda_n}+\lambda_n(\text{Re}\langle\partial_{\mu}n| \partial_\nu n\rangle+\mathcal{A}_{n\mu}\mathcal{A}_{n\nu})\right]$ \\\hline
  Imaginary part of QGT & Berry curvature $\frac{\mi}{2}\mathcal{F}$ & Weighted sum of Berry curvature $\frac{\mi}{2}\sum_n\lambda_n\mathcal{F}_n$\\\hline
  \end{tabular}
  \end{table}

Since all contribution from the phase factors has been eliminated by the projection $\pi$, $\dif s^2 (\mathcal{D}^N_N)$ or the Sj\"oqvist distance must remain  invariant under any local U$^N(1)$ gauge transformation. The corresponding metric is given by Eq.(\ref{Sm0}), which is also U$^N(1)$-invariant. As we stated before, it is also referred to as the U$^N(1)$ QGT. According to the decomposition of distances, the U$^N(1)$ QGT can also be decomposed into
\begin{align}\label{Sm1}
Q^\text{S}_{\mu\nu}=g^\text{FR}_{\mu\nu}+g^\text{FS}_{\mu\nu}-\mi\Omega_{\mu\nu},
\end{align}
where
\begin{align}
g^\text{FR}_{\mu\nu}=\sum_n\frac{\partial_\mu\lambda_n\partial_\nu\lambda_n}{4\lambda_n}
\end{align}
is the Fisher-Rao metric,
\begin{align}
 g^\text{FS}_{\mu\nu}=\sum_n\lambda_n g^{\text{FS}}_{n\mu\nu}=\sum_n\lambda_n(\text{Re}\langle\partial_{\mu}n| \partial_\nu n\rangle+\mathcal{A}_{n\mu}\mathcal{A}_{n\nu})
\end{align}
is the weighted sum of the Fubini-Study metrics for each spectral ray, and the negative imaginary part
\begin{align}\label{iofSm}
\Omega_{\mu\nu}=\frac{\mi}{2}\sum_n\lambda_n\left(\langle\partial_\mu n|\partial_\nu n\rangle-\langle\partial_\nu n|\partial_\mu n\rangle\right)\equiv \frac{\mi}{2}\sum_n\lambda_n\mathcal{F}_{n\mu\nu}
\end{align}
is half of the weighted sum of the Berry curvature for each spectral ray. Comparing to discussions in Sec.\ref{I.B}, the formalism developed here indeed serves as a suitable generalization to that of pure states. We summarize our main results by comparing the key points between pure and mixed states in Table \ref{table1}.

Since $\lambda_n$ generically depends on $\mathbf{R}$, $\Omega_{\mu\nu}$ is not necessarily the gauge potential associated with $\mathcal{A}_{\mathcal{D}^N_N}$. In other words, $\Omega_{\mu\nu}=\frac{\mi}{2}\left[\partial_\mu(\sum_n\lambda_n\mathcal{A}_{n\nu})-\partial_\nu(\sum_n\lambda_n\mathcal{A}_{n\mu})\right]$ is not true in general. Nevertheless, $\Omega_{\mu\nu}$ is a gauge-invariant quantity. We introduce the 2-form $\Omega=\frac{1}{2}\Omega_{\mu\nu}\dif R^\mu\wedge \dif R^\nu=\frac{\mi}{2}\sum_n\lambda_n \mathcal{F}_n$, where $\mathcal{F}_n=\frac{1}{2}\mathcal{F}_{n\mu\nu}\dif R^\mu\wedge \dif R^\nu$ is is the Berry curvature 2-form for the $n$-th level. The integral of $\Omega$ on a parameter surface $\Sigma$ is also gauge-invariant. Let $C$ be the boundary of
$\Sigma$. We further define a gauge-invariant quantity
\begin{align}\label{aofSm}
\theta_g(C)=\int_\Sigma\Omega=\frac{\mi}{2}\sum_n\int_\Sigma\lambda_n\mathcal{F}_{n}.
\end{align}
If all $\lambda_n$ are constant on $\Sigma$, then we have
\begin{align}\label{aofSm3}
\theta_g(C)=\frac{\mi}{2}\sum_n\lambda_n\int_\Sigma\mathcal{F}_{n}=\frac{1}{2}\sum_n\lambda_n\theta_{\text{B}n}(C),
\end{align}
which is half the weighted sum of all Berry phases.

\subsection{Fundamental inequalities about the U$^N(1)$ QGT}

It is known that the pure-state QGT, as described in Eq.(\ref{qd4}), satisfies a fundamental inequality
\begin{align}
Q_{\mu\mu}Q_{\nu\nu}\ge|Q_{\mu\nu}|^2,
\end{align}
which has intriguing applications in condensed matter physics, particularly in two-dimensional models \cite{PhysRevB.90.165139,PhysRevB.104.045103}. Here we show that the U$^N(1)$ QGT for mixed states holds the same result:
\begin{align}\label{Sm0iqn}
Q^\text{S}_{\mu\mu}Q^\text{S}_{\nu\nu}\ge|Q^\text{S}_{\mu\nu}|^2.
\end{align}
Let $g^\text{FR}_{n\mu\nu}=\frac{\partial_\mu\lambda_n\partial_\nu\lambda_n}{4\lambda_n}$ and $Q^n_{\mu\nu}=g^\text{FS}_{n\mu\nu}-\frac{\mi}{2}\mathcal{F}_{n\mu\nu}$ represent the Fisher-Rao metric and the pure-state QGT for the
$n$-th eigenstate of $\rho$, respectively. Then, $Q^\text{S}_{\mu\nu}=\sum_n\left(g^\text{FR}_{n\mu\nu}+\lambda_nQ^n_{\mu\nu}\right)$. We first prove the follow inequalities:
\begin{align}
\sum_ng^\text{FR}_{n\mu\mu}\sum_mg^\text{FR}_{m\nu\nu}&\ge|\sum_ng^\text{FR}_{n\mu\nu}|^2,\label{Sm0iqn2} \\ \sum_n\lambda_nQ^n_{\mu\mu}\sum_m\lambda_mQ^m_{\nu\nu}&\ge|\sum_n\lambda_nQ^n_{\mu\nu}|^2.\label{Sm0iqn3}
\end{align}
Noting $g^\text{FR}_{n\mu\mu}=\left(\frac{\partial_\mu\lambda_n}{2\sqrt{\lambda_n}}\right)^2$ and applying the Cauchy-Schwarz inequality
 \begin{align}\label{C-Siqn1}
\sum_{mn}\left(\frac{\partial_\mu\lambda_n}{2\sqrt{\lambda_n}}\right)^2\left(\frac{\partial_\nu\lambda_m}{2\sqrt{\lambda_m}}\right)^2\ge
\left(\sum_{n}\frac{\partial_\mu\lambda_n\partial_\nu\lambda_n}{4\lambda_n}\right)^2,
\end{align}
we find that Inequality (\ref{Sm0iqn2}) is clearly valid. To prove Inequality (\ref{Sm0iqn3}), we introduce the projection operator $P_n=|n\rangle\langle n|$, leading to $Q^n_{\mu\nu}=\langle\partial_\mu n|(1-P_n)|\partial_\nu n\rangle$. Next, let
\begin{align}
|\alpha\rangle
=\begin{pmatrix}
\sqrt{\lambda_0}(1-P_0 )|\partial_\mu0 \rangle\\\sqrt{\lambda_1}(1-P_1 )|\partial_\mu1 \rangle\\\vdots\\\sqrt{\lambda_{N-1}}(1-P_{N-1} )|\partial_\mu (N-1) \rangle
\end{pmatrix},\quad
|\beta\rangle
=\begin{pmatrix}
\sqrt{\lambda_0}(1-P_0 )|\partial_\nu0 \rangle\\\sqrt{\lambda_1}(1-P_1 )|\partial_\nu1 \rangle\\\vdots\\\sqrt{\lambda_{N-1}}(1-P_{N-1} )|\partial_\nu (N-1) \rangle
\end{pmatrix}.
\end{align}
Using $(1-P_n)^2=1-P_n$, we obtain $\langle \alpha|\alpha\rangle=\sum_n\lambda_nQ^n_{\mu\mu}$, $\langle \beta|\beta\rangle=\sum_m\lambda_mQ^m_{\nu\nu}$ and $\langle \alpha|\beta\rangle=\sum_n\lambda_nQ^n_{\mu\nu}$. By again applying the Cauchy-Schwarz inequality $\langle\alpha|\alpha\rangle\langle\beta|\beta\rangle\ge |\langle\alpha|\beta\rangle|^2$, we can deduce Inequality (\ref{Sm0iqn3}). The same proof process can also yield the following result (Taking $N=1$ and $\lambda_n=1$): \begin{align}\label{Qgiene}
Q^n_{\mu\mu}Q^n_{\nu\nu}\ge|Q^n_{\mu\nu}|^2.
\end{align}
The two sides of Inequality (\ref{Sm0iqn}) are respectively expanded as
\begin{align}\label{Sm0iqnb}
Q^\text{S}_{\mu\mu}Q^\text{S}_{\nu\nu}
=&\sum_{nm}\left(g^\text{FR}_{n\mu\mu}g^\text{FR}_{m\nu\nu}+\lambda_nQ^n_{\mu\mu}\lambda_mQ^m_{\nu\nu}\right)+\sum_{nm}g^\text{FR}_{n\mu\mu}\lambda_m\left(Q^m_{\mu\mu}+Q^m_{\nu\nu}\right),\notag\\
|Q^\text{S}_{\mu\nu}|^2=&|\sum_ng^\text{FR}_{n\mu\nu}|^2+|\sum_n\lambda_nQ^n_{\mu\nu}|^2+\sum_{nm}\lambda_mg^\text{FR}_{n\mu\nu}\left(Q^m_{\mu\nu}+\bar{Q}^m_{\mu\nu}\right),
\end{align}
where we have interchanged the indices $n$ and $m$ in the cross terms.
Inequalities (\ref{Sm0iqn2}) and (\ref{Sm0iqn3}) imply
\begin{align}\label{Sm0iqnc}
\sum_{nm}\left(g^\text{FR}_{n\mu\mu}g^\text{FR}_{m\nu\nu}+\lambda_nQ^n_{\mu\mu}\lambda_mQ^m_{\nu\nu}\right)\ge|\sum_ng^\text{FR}_{n\mu\nu}|^2+|\sum_n\lambda_nQ^n_{\mu\nu}|^2.
\end{align}
Next, applying the facts that $g^\text{FR}_{n\mu\mu}\ge 0$, $\lambda_n\ge 0$ and $g^\text{FR}_{n\mu\mu}g^\text{FR}_{n\nu\nu}=\left(g^\text{FR}_{n\mu\nu}\right)^2$, and Inequality (\ref{Qgiene}), it can be found that
\begin{align}\label{Sm0iqnd}
\sum_{nm}\lambda_m\left(g^\text{FR}_{n\mu\mu}Q^m_{\nu\nu}+g^\text{FR}_{n\nu\nu}Q^m_{\mu\mu}\right)\ge2\sum_{nm}\lambda_m\sqrt{g^\text{FR}_{n\mu\mu}g^\text{FR}_{n\nu\nu}Q^m_{\mu\mu}Q^m_{\nu\nu}}\ge \sum_{nm}g^\text{FR}_{n\mu\nu}\lambda_m2|Q^m_{\mu\nu}|\ge\sum_{nm}g^\text{FR}_{n\mu\nu}\lambda_m(Q^m_{\mu\nu}+\bar{Q}^m_{\mu\nu}).
\end{align}
Finally, this inequality, together with Inequality (\ref{Sm0iqnc}), can jointly deduce Inequality (\ref{Sm0iqn}). In Ref.\cite{PhysRevB.110.035144}, we pointed out that the real part of the U($N$) QGT is the Bures metric. In fact, Bures metric also respect a similar inequality. We present the proof in the Appendix \ref{appa}.

If the parameter space is two dimensional, i.e. $\mathbf{R}=(R^1,R^2)$, the U$^N(1)$ QGT can be expressed as a $2\times2$ matrix and the nontrivial result of Inequality (\ref{Sm0iqn}) is
\begin{align}\label{Sm0iqng}
Q^\text{S}_{11}Q^\text{S}_{22}\ge|Q^\text{S}_{12}|^2.
\end{align}
Let $g^\text{S}_{\mu\nu}$ be the real part of $Q^\text{S}_{\mu\nu}$. For convenience, we also introduce $\mathcal{F}_{\mu\nu}=-2\text{Im}Q^\text{S}_{\mu\nu}=\sum_n\lambda_n\mathcal{F}_{n\mu\nu}$. The U$^N(1)$ QGT can be written as
\begin{align}\label{SJQGTmr}
Q^\text{S} =\begin{pmatrix}g^\text{s}_{11}  & g^\text{S}_{12} -\frac{\mi}{2}\mathcal{F}_{12} \\
g^\text{S}_{21} -\frac{\mi}{2}\mathcal{F}_{21}  & g^\text{S}_{22}
\end{pmatrix}=g^\text{S}-\frac{\mi}{2}\mathcal{F}_{12}\begin{pmatrix} 0 & 1\\ -1 & 0\end{pmatrix}.
\end{align}
The Inequality (\ref{Sm0iqng}) implies
\begin{align}\label{Sm0iqng2}
\det(g^\text{S})=g^\text{S}_{11}g^\text{S}_{22}-(g^\text{S}_{12})^2\ge\frac{1}{4}|\mathcal{F}_{12}|^2,\text{ or }\sqrt{ \det(g^\text{S})}\ge\frac{|\mathcal{F}_{12}|}{2}.
\end{align}
This is a direct generalization to the results for pure-states \cite{PhysRevB.104.045103}. Define
\begin{align}
V^\text{S}_g=\int\dif^2\mathbf{R}\sqrt{\det(g^\text{S})},
\end{align}
which is the quantum volume of the parameter space. Inequality (\ref{Sm0iqng2}) yields
\begin{align}
V^\text{S}_g\ge\int\dif^2\mathbf{R}\frac{|\mathcal{F}_{12}|}{2}\ge\left|\frac{\mi}{2}\int\dif^2\mathbf{R}\sum_n\lambda_n\mathcal{F}_{n12}\right|=|\theta_g|.
\end{align}
This builds a relation between $V^\text{S}_g$ and $\theta_g$.

\subsection{Examples of U$^N(1)$ QGT}

\subsubsection{Bosonic coherent state}

We first calculate the U$^N(1)$ QGT of bosonic coherent states, which may be constructed from bosonic harmonic oscillators~\cite{Swanson_Book,Scully_Book}. The Hamiltonian of a single harmonic oscillator is given by $\hat{H}=\hbar\omega(a^\dagger a+\frac{1}{2})$,
where $a$ and $a^\dag$ are the annihilation and creation operators, respectively, satisfying $[a,a^\dag]=1$. The energy levels of system satisfy  $\hat{H}|n\rangle=\hbar\omega(n+\frac{1}{2})|n\rangle$ with $n=0,1,2,\cdots$. The coherent state is defined by operating the translation operator on the ground state: $|z\rangle=D(z)|0\rangle\equiv\me^{za^\dagger-\bar{z}a}|0\rangle$. This can be generalized to exited state: $|n,z\rangle=D(z)|n\rangle$, $n\ge1$. The density matrix is obtained via the transformation $D(z)$: $\rho(z)=\frac{1}{Z}\me^{-\beta\hat{H}(z)}=D(z)\rho(0)D^\dag(z)$, where $\rho(0)=\frac{1}{Z}\me^{-\beta\hat{H}}$ with $\beta$ being the inverse temperature. Since $\rho(0)$ shares the same eigenstates with $\hat{H}$, the eigenstates of $\rho(z)$ are $|n,z\rangle$ with $n=0,1,2,\cdots$. Note the eigenvalues $\lambda_n=\frac{1}{Z}\me^{-\beta\hbar\omega (n+\frac{1}{2})}$ are constants, then the Sj\"oqvist distance is
\begin{align}
\dif s^2_\text{S}=&\sum_n\lambda_n\dif\langle n ,z|(1-|n,z\rangle\langle n,z|)\dif |n,z\rangle\notag\\
=&\sum_n\lambda_n\left[ \langle n |\dif D^\dag(z)\dif D(z)|n\rangle-\langle n |\dif D^\dag(z)D(z)|n\rangle\langle n|D^\dag(z)\dif D(z) |n\rangle\right].
\end{align}
A straightforward evaluation shows
\begin{align}
	&\langle n |\dif D^\dag(z)\dif D(z)|n\rangle=\left(2n+1+\frac{{|z|}^2}{2}\right)\dif z \dif \bar{z}-\frac{{z}^2}{4} \dif \bar{z}^2 -\frac{{\bar{z}}^2}{4} \dif z^2,\notag\\
&\langle n |\dif D^\dag(z)D(z)|n\rangle\langle n|D^\dag(z)\dif D(z) |n\rangle=- \frac{(\bar{z} \dif  z-z \dif \bar{z})^2 }{4} .
\end{align}
This leads to
\begin{align}
	\dif s^2_\text{S}
	=&\sum_n (2n+1)\lambda_n \dif z \dif \bar{z}= \text{coth}\left(\frac{\beta\hbar\omega}{2}\right) \dif z\dif \bar{z}=\text{coth}\left(\frac{\beta\hbar\omega}{2}\right)(\dif x^2+\dif y^2).
\end{align}
In the zero temperature limit, it reduces to the Fubini-Study distance\cite{QGTCMP80}: $\dif s^2_\text{S}=\dif x^2+\dif y^2$. This is reasonable since the  contributions from all excited states fade away as $\beta\rightarrow +\infty$. At the infinite temperature,
\begin{align}
\lim_{\beta\rightarrow 0}\dif s^2_\text{S}=(\dif x^2+\dif y^2)\lim_{\beta\rightarrow 0}\text{coth}\left(\frac{\beta\hbar\omega}{2}\right)\rightarrow +\infty.
\end{align}
Even the Sj\"oqvist distance between adjacent coherent states diverges. This may be due to the existence of an infinite number of energy levels, each of which carries the same weight at infinite temperature, resulting in a divergent total contribution from the Fubini-Study distances of all energy levels.

The real part of the U$^N(1)$ QGT can be immediately obtained from the expression of $\dif s^2_\text{S}$. According to Eq.(\ref{iofSm}), in terms of 2-form, the negative imaginary part is
 \begin{align}
\Omega=&\frac{\mi}{2}\sum_n\lambda_n\dif\langle n,z|\wedge\dif|n,z\rangle=\frac{\mi}{2}\sum_n\lambda_n\langle n |\dif D^\dag(z)\wedge\dif D(z)|n\rangle\notag\\
=&\frac{\mi}{2}\sum_n\lambda_n\langle n |\left[\left(a^\dag+\frac{\bar{z}}{2}\right)\left(a+\frac{z}{2}\right)-\left(a+\frac{z}{2}\right)\left(a^\dag+\frac{\bar{z}}{2}\right)\right]|n\rangle\dif z\wedge\dif \bar{z}\notag\\
=&-\frac{\mi}{2}\dif z\wedge\dif \bar{z}=-\dif x\wedge \dif y.
\end{align}

\subsubsection{Fermionic coherent state}

The Hamiltonian of a fermionic harmonic oscillator is $\hat{H}=\frac{\hbar\omega}{2}[b^\dag,b]=\hbar\omega\left(b^\dag b-\frac{1}{2}\right)$, where the fermionic operators $b$ and $b^\dag$ satisfy the algebra $\{b,b^\dag\}=1$ \cite{Swanson_Book,Kleinert_Book}. Similar to its bosonic counterpart, the fermionic coherent state is also constructed by applying a translation to the vacuum: $|\xi\rangle=D(\xi)|0\rangle\equiv\me^{b^\dagger \xi-\bar{\xi}b}|0\rangle$, where $\xi$ is a Grassmann number and anticommutes with any fermionic operator. The corresponding density matrix is
\begin{align}\label{rhozf}
    \rho(\xi )=\frac{1}{Z}\me^{-\beta D(\xi )\hat{H}D^\dag(\xi )}=D(\xi )\rho(0)D^\dag(\xi ),
\end{align}
where $\rho(0)=\frac{\me^{-\beta \hat{H}}}{Z}$ with the partition function $Z=\me^{\frac{1}{2}\beta\hbar\omega}+\me^{-\frac{1}{2}\beta\hbar\omega}=2\cosh\frac{\beta\hbar\omega}{2}$. Using the algebra $(b^\dag b)^2=b^\dag b$, it can be found that
\begin{align}\label{rxi}
\rho(\xi)
&=\frac{1}{1+\me^{-\beta\hbar\omega}}-\tanh\left(\frac{\beta\hbar\omega}{2}\right)(b^\dagger-\bar{\xi})(b-\xi).
\end{align}
The eigenvalues of $\rho(\xi)$ are independent of $\xi$, then the Sj\"oqvist distance is
\begin{align}
\dif s^2_\text{S}
=&\sum_{n=0,1}\lambda_n \langle n |\dif D^\dag(\xi)[1-D(\xi)|n\rangle\langle n|D^\dag(\xi)]\dif D(\xi) |n\rangle.
\end{align}
Applying
\begin{align}
	\mathrm{d} D(\xi)=&\left(b^{\dagger}-\frac{1}{2} \bar{\xi}\right) D(\xi) \mathrm{d} \xi+D(\xi)\left(b+\frac{1}{2} \xi\right) \mathrm{d} \bar{\xi} \notag\\
	\mathrm{d} D^{\dagger}(\xi)=&-D^{\dagger}(\xi)\left(b-\frac{1}{2} \xi \right)  \mathrm{d} \bar{\xi} -\left(b^{\dagger}+\frac{1}{2} \bar{\xi} \right) D^{\dagger}(\xi) \mathrm{d} \xi,
\end{align}
we obtain
\begin{align}
\dif s^2_\text{S}= \lambda_0 \mathrm{d} \xi \mathrm{d} \bar{\xi}-\lambda_1 \mathrm{d} \bar{\xi}\mathrm{d} \xi =\tanh\left(\frac{\beta \hbar \omega}{2}\right) \mathrm{d} \bar{\xi} \mathrm{d} \xi .
\end{align}
In the zero temperature limit, it reduces to $\dif s^2_\text{S}=\mathrm{d} \bar{\xi} \mathrm{d} \xi $. At the infinite temperature, $\dif s^2_\text{S}\rightarrow 0$, which sharply contrasts with the results for bosonic coherent states. In terms of 2-form, the negative imaginary part is
 \begin{align}
\Omega
=&\frac{\mi}{2}\sum_n\lambda_n\langle n |\dif D^\dag(\xi)\wedge\dif D(\xi)|n\rangle\notag\\
=&\frac{\mi}{2}\sum_n\lambda_n\left[\langle n |\left( bb^{\dagger}+\frac{\xi\bar{\xi}}{4} \right)\mathrm{d} \bar{\xi}\wedge \mathrm{d} \xi |n\rangle    + \langle n |  \left( b^{\dagger}b +\frac{ \bar{\xi}\xi }{4}\right) \mathrm{d} \xi \wedge\mathrm{d} \bar{\xi} |n\rangle\right] \notag\\
=&\dif \xi\wedge\dif \bar{\xi}.
\end{align}

\section{Conclusion}\label{III}

In conclusion, we have established a comprehensive mathematical framework for the U$^N(1)$ QGT applicable to mixed states, significantly enhancing our understanding of quantum state geometry. By employing the U$^N(1)$ principal bundle, we introduced the Ehresmann connection, explored the bundle's fibration, and derived a gauge-invariant metric, along with a Pythagorean-like distance decomposition that highlights the geometric relationships inherent in mixed-state manifolds. Our comparative analysis with the U(1) principal bundle highlights the distinct properties of the U$^N(1)$ QGT, emphasizing its Riemannian characteristics and the inclusion of a nonzero imaginary part, which is crucial for capturing the complexity of mixed states.
Furthermore, the proof of a fundamental inequality for the U$^N(1)$ QGT reinforces the theoretical foundations. Overall, the work presented here paves a step towards a deeper understanding of the geometric and topological aspects of mixed quantum states.

\section{Conflict of interest}
The authors declared that they have no conflicts of interest to this work.

\section{Data availability}
The authors confirm that the data supporting the findings of this study are available within the article.

\section{Acknowledgments}
H.G. was supported by the Innovation Program for Quantum
Science and Technology (Grant No. 2021ZD0301904) and the National Natural Science
Foundation of China (Grant No. 12074064). X. Y. H. was supported by the Jiangsu Funding Program for Excellent Postdoctoral Talent (Grant No. 2023ZB611).

\appendix
\section{Fundamental inequality for Bures metric}\label{appa}
The Bures metric is given by \cite{HubnerPLA92}
 \begin{align}\label{Bmetric1}
 g^\text{B}_{\mu\nu}=\frac{1}{2}\sum_{ij}\frac{\langle i|\partial_\mu\rho|j\rangle\langle j|\partial_\nu\rho|i\rangle}{\lambda_i+\lambda_j}.
\end{align}
We prove that it also satisfies the inequality
\begin{align}\label{Bien}
 g^\text{B}_{\mu\mu} g^\text{B}_{\nu\nu}\ge\left( g^\text{B}_{\mu\nu}\right)^2.
\end{align}

Proof: Using $\partial_\mu\rho=\sum_n\left[\partial_\mu\lambda_n|n\rangle\langle n|+\lambda_n(|\partial_\mu n\rangle\langle n|+| n\rangle\langle \partial_\mu n|)\right]$, we obtain  \begin{align}
\langle m| \partial_\mu\rho|n\rangle= \partial_\mu\lambda_n\delta_{mn}+(\lambda_n-\lambda_m)\langle m|\partial_\mu n\rangle .
\end{align}
Substitute it into Eq.(\ref{Bmetric1}), the Bures metric can be reexpressed as
 \begin{align}\label{gbmn}
g^\text{B}_{\mu\nu}=\frac{1}{4}\sum_n\frac{\partial_\mu\lambda_n\partial_\nu\lambda_n}{\lambda_n}
+\frac{1}{2}\sum_{nm}\frac{(\lambda_n-\lambda_m)^2}{\lambda_n+\lambda_m}\mi\langle m|\partial_\mu n\rangle\mi\langle n|\partial_\nu m\rangle,
\end{align}
where in the second item, we have introduced the factor ``i" to ensure that $\mi\langle m|\partial_\mu n\rangle$ and $\mi\langle n|\partial_\nu m\rangle$ are both real numbers. For convenience, we let $A_{mn}=\frac{1}{2}\frac{(\lambda_n-\lambda_m)^2}{\lambda_n+\lambda_m}$, which satisfies $A_{mn}\ge0$ and $A_{mn}=A_{nm}$. According to Eq.(\ref{Bmetric1}), we have
 \begin{align}
g^\text{B}_{\mu\mu}=\sum_n\left(\frac{\partial_\mu\lambda_n}{2\sqrt{\lambda_n}}\right)^2+\sum_{mn}A_{mn}|\langle m|\partial_\mu n\rangle|^2.
\end{align}
Thus,
 \begin{align}
g^\text{B}_{\mu\mu}g^\text{B}_{\nu\nu}=&\sum_{mn}\left(\frac{\partial_\mu\lambda_n}{2\sqrt{\lambda_n}}\right)^2\left(\frac{\partial_\nu\lambda_m}{2\sqrt{\lambda_m}}\right)^2
+\sum_{mnkl}A_{mn}A_{kl}|\langle m|\partial_\mu n\rangle|^2|\langle k|\partial_\nu l\rangle|^2\notag\\
+&\sum_{mnk}\left[\left(\frac{\partial_\mu\lambda_k}{2\sqrt{\lambda_k}}\right)^2A_{mn}|\langle m|\partial_\nu n\rangle|^2
+\left(\frac{\partial_\nu\lambda_k}{2\sqrt{\lambda_k}}\right)^2A_{mn}|\langle m|\partial_\mu n\rangle|^2\right].
\end{align}
Applying Cauchy-Schwarz inequality, we obtain
 \begin{align}
\sum_{mnkl}A_{mn}A_{kl}|\langle m|\partial_\mu n\rangle|^2|\langle k|\partial_\nu l\rangle|^2\ge
\sum_{mn}A^2_{mn}\left|\langle m|\partial_\mu n\rangle \langle n|\partial_\nu m\rangle\right|^2.
\end{align}
Next, applying Inequality (\ref{C-Siqn1}) and the mean value inequality, it can be found
 \begin{align}
\left(\frac{\partial_\mu\lambda_k}{2\sqrt{\lambda_k}}\right)^2A_{mn}|\langle m|\partial_\nu n\rangle|^2
+\left(\frac{\partial_\nu\lambda_k}{2\sqrt{\lambda_k}}\right)^2A_{mn}|\langle m|\partial_\mu n\rangle|^2\ge\sum_k\frac{\partial_\mu\lambda_k\partial_\nu\lambda_k}{2\lambda_k}A_{mn}|\langle m|\partial_\nu n\rangle||\langle m|\partial_\mu n\rangle|.
\end{align}
Combining the above results, we can finally prove Inequality (\ref{Bien}).

\bibliographystyle{apsrev}

\begin{thebibliography}{42}
\expandafter\ifx\csname natexlab\endcsname\relax\def\natexlab#1{#1}\fi
\expandafter\ifx\csname bibnamefont\endcsname\relax
  \def\bibnamefont#1{#1}\fi
\expandafter\ifx\csname bibfnamefont\endcsname\relax
  \def\bibfnamefont#1{#1}\fi
\expandafter\ifx\csname citenamefont\endcsname\relax
  \def\citenamefont#1{#1}\fi
\expandafter\ifx\csname url\endcsname\relax
  \def\url#1{\texttt{#1}}\fi
\expandafter\ifx\csname urlprefix\endcsname\relax\def\urlprefix{URL }\fi
\providecommand{\bibinfo}[2]{#2}
\providecommand{\eprint}[2][]{\url{#2}}

\bibitem[{\citenamefont{Provost and Vallee}(1980)}]{QGTCMP80}
\bibinfo{author}{\bibfnamefont{J.~P.} \bibnamefont{Provost}} \bibnamefont{and}
  \bibinfo{author}{\bibfnamefont{G.}~\bibnamefont{Vallee}},
  \bibinfo{journal}{Commun. Math. Phys.} \textbf{\bibinfo{volume}{76}},
  \bibinfo{pages}{289} (\bibinfo{year}{1980}).

\bibitem[{\citenamefont{Cheng}(2010)}]{QGT10}
\bibinfo{author}{\bibfnamefont{R.}~\bibnamefont{Cheng}},
  \emph{\bibinfo{title}{Quantum geometric tensor (fubini-study metric) in
  simple quantum system: A pedagogical introduction}} (\bibinfo{year}{2010}),
  \bibinfo{note}{arXiv:1012.1337}.

\bibitem[{\citenamefont{Braunstein and Caves}(1994)}]{PhysRevLett.72.3439}
\bibinfo{author}{\bibfnamefont{S.~L.} \bibnamefont{Braunstein}}
  \bibnamefont{and} \bibinfo{author}{\bibfnamefont{C.~M.} \bibnamefont{Caves}},
  \bibinfo{journal}{Phys. Rev. Lett.} \textbf{\bibinfo{volume}{72}},
  \bibinfo{pages}{3439} (\bibinfo{year}{1994}),
  \urlprefix\url{https://link.aps.org/doi/10.1103/PhysRevLett.72.3439}.

\bibitem[{\citenamefont{Kolodrubetz et~al.}(2017)\citenamefont{Kolodrubetz,
  Sels, Mehta, and Polkovnikov}}]{KOLODRUBETZ20171}
\bibinfo{author}{\bibfnamefont{M.}~\bibnamefont{Kolodrubetz}},
  \bibinfo{author}{\bibfnamefont{D.}~\bibnamefont{Sels}},
  \bibinfo{author}{\bibfnamefont{P.}~\bibnamefont{Mehta}}, \bibnamefont{and}
  \bibinfo{author}{\bibfnamefont{A.}~\bibnamefont{Polkovnikov}},
  \bibinfo{journal}{Physics Reports} \textbf{\bibinfo{volume}{697}},
  \bibinfo{pages}{1} (\bibinfo{year}{2017}), ISSN \bibinfo{issn}{0370-1573},
  \urlprefix\url{https://www.sciencedirect.com/science/article/pii/S0370157317301989}.

\bibitem[{\citenamefont{Amari}(2016)}]{IG_Book}
\bibinfo{author}{\bibfnamefont{S.-i.} \bibnamefont{Amari}},
  \emph{\bibinfo{title}{Information Geometry and Its Applications}}
  (\bibinfo{publisher}{Springer Japan}, \bibinfo{address}{Tokyo},
  \bibinfo{year}{2016}).

\bibitem[{\citenamefont{Ercolessi et~al.}(2024)\citenamefont{Ercolessi,
  Fioresi, and Weber}}]{doi:10.1142/S0219887824400115}
\bibinfo{author}{\bibfnamefont{E.}~\bibnamefont{Ercolessi}},
  \bibinfo{author}{\bibfnamefont{R.}~\bibnamefont{Fioresi}}, \bibnamefont{and}
  \bibinfo{author}{\bibfnamefont{T.}~\bibnamefont{Weber}},
  \bibinfo{journal}{International Journal of Geometric Methods in Modern
  Physics} \textbf{\bibinfo{volume}{21}}, \bibinfo{pages}{2440011}
  (\bibinfo{year}{2024}).

\bibitem[{\citenamefont{Ozawa and Mera}(2021)}]{PhysRevB.104.045103}
\bibinfo{author}{\bibfnamefont{T.}~\bibnamefont{Ozawa}} \bibnamefont{and}
  \bibinfo{author}{\bibfnamefont{B.}~\bibnamefont{Mera}},
  \bibinfo{journal}{Phys. Rev. B} \textbf{\bibinfo{volume}{104}},
  \bibinfo{pages}{045103} (\bibinfo{year}{2021}),
  \urlprefix\url{https://link.aps.org/doi/10.1103/PhysRevB.104.045103}.

\bibitem[{\citenamefont{Ozawa}(2018)}]{PhysRevB.97.041108}
\bibinfo{author}{\bibfnamefont{T.}~\bibnamefont{Ozawa}},
  \bibinfo{journal}{Phys. Rev. B} \textbf{\bibinfo{volume}{97}},
  \bibinfo{pages}{041108} (\bibinfo{year}{2018}),
  \urlprefix\url{https://link.aps.org/doi/10.1103/PhysRevB.97.041108}.

\bibitem[{\citenamefont{Klees et~al.}(2021)\citenamefont{Klees, Cuevas, Belzig,
  and Rastelli}}]{PhysRevB.103.014516}
\bibinfo{author}{\bibfnamefont{R.~L.} \bibnamefont{Klees}},
  \bibinfo{author}{\bibfnamefont{J.~C.} \bibnamefont{Cuevas}},
  \bibinfo{author}{\bibfnamefont{W.}~\bibnamefont{Belzig}}, \bibnamefont{and}
  \bibinfo{author}{\bibfnamefont{G.}~\bibnamefont{Rastelli}},
  \bibinfo{journal}{Phys. Rev. B} \textbf{\bibinfo{volume}{103}},
  \bibinfo{pages}{014516} (\bibinfo{year}{2021}),
  \urlprefix\url{https://link.aps.org/doi/10.1103/PhysRevB.103.014516}.

\bibitem[{\citenamefont{Porlles and Chen}(2023)}]{PhysRevB.108.094508}
\bibinfo{author}{\bibfnamefont{D.}~\bibnamefont{Porlles}} \bibnamefont{and}
  \bibinfo{author}{\bibfnamefont{W.}~\bibnamefont{Chen}},
  \bibinfo{journal}{Phys. Rev. B} \textbf{\bibinfo{volume}{108}},
  \bibinfo{pages}{094508} (\bibinfo{year}{2023}),
  \urlprefix\url{https://link.aps.org/doi/10.1103/PhysRevB.108.094508}.

\bibitem[{\citenamefont{Neupert et~al.}(2013)\citenamefont{Neupert, Chamon, and
  Mudry}}]{PhysRevB.87.245103}
\bibinfo{author}{\bibfnamefont{T.}~\bibnamefont{Neupert}},
  \bibinfo{author}{\bibfnamefont{C.}~\bibnamefont{Chamon}}, \bibnamefont{and}
  \bibinfo{author}{\bibfnamefont{C.}~\bibnamefont{Mudry}},
  \bibinfo{journal}{Phys. Rev. B} \textbf{\bibinfo{volume}{87}},
  \bibinfo{pages}{245103} (\bibinfo{year}{2013}),
  \urlprefix\url{https://link.aps.org/doi/10.1103/PhysRevB.87.245103}.

\bibitem[{\citenamefont{Ahn et~al.}(2020)\citenamefont{Ahn, Guo, and
  Nagaosa}}]{PhysRevX.10.041041}
\bibinfo{author}{\bibfnamefont{J.}~\bibnamefont{Ahn}},
  \bibinfo{author}{\bibfnamefont{G.-Y.} \bibnamefont{Guo}}, \bibnamefont{and}
  \bibinfo{author}{\bibfnamefont{N.}~\bibnamefont{Nagaosa}},
  \bibinfo{journal}{Phys. Rev. X} \textbf{\bibinfo{volume}{10}},
  \bibinfo{pages}{041041} (\bibinfo{year}{2020}),
  \urlprefix\url{https://link.aps.org/doi/10.1103/PhysRevX.10.041041}.

\bibitem[{\citenamefont{Bhattacharya and
  Black-Schaffer}(2024)}]{Bhattacharya24}
\bibinfo{author}{\bibfnamefont{A.}~\bibnamefont{Bhattacharya}}
  \bibnamefont{and} \bibinfo{author}{\bibfnamefont{A.~M.}
  \bibnamefont{Black-Schaffer}}, \emph{\bibinfo{title}{Electric field induced
  second-order anomalous hall transport in unconventional rashba system}}
  (\bibinfo{year}{2024}), \bibinfo{note}{arXiv:2408.15840},
  \urlprefix\url{https://arxiv.org/pdf/2408.15840}.

\bibitem[{\citenamefont{Zhang et~al.}(2019)\citenamefont{Zhang, Wang, and
  Gong}}]{PhysRevA.99.042104}
\bibinfo{author}{\bibfnamefont{D.-J.} \bibnamefont{Zhang}},
  \bibinfo{author}{\bibfnamefont{Q.-h.} \bibnamefont{Wang}}, \bibnamefont{and}
  \bibinfo{author}{\bibfnamefont{J.}~\bibnamefont{Gong}},
  \bibinfo{journal}{Phys. Rev. A} \textbf{\bibinfo{volume}{99}},
  \bibinfo{pages}{042104} (\bibinfo{year}{2019}),
  \urlprefix\url{https://link.aps.org/doi/10.1103/PhysRevA.99.042104}.

\bibitem[{\citenamefont{Pi\'echon et~al.}(2016)\citenamefont{Pi\'echon, Raoux,
  Fuchs, and Montambaux}}]{PhysRevB.94.134423}
\bibinfo{author}{\bibfnamefont{F.}~\bibnamefont{Pi\'echon}},
  \bibinfo{author}{\bibfnamefont{A.}~\bibnamefont{Raoux}},
  \bibinfo{author}{\bibfnamefont{J.-N.} \bibnamefont{Fuchs}}, \bibnamefont{and}
  \bibinfo{author}{\bibfnamefont{G.}~\bibnamefont{Montambaux}},
  \bibinfo{journal}{Phys. Rev. B} \textbf{\bibinfo{volume}{94}},
  \bibinfo{pages}{134423} (\bibinfo{year}{2016}),
  \urlprefix\url{https://link.aps.org/doi/10.1103/PhysRevB.94.134423}.

\bibitem[{\citenamefont{Gao et~al.}(2015)\citenamefont{Gao, Yang, and
  Niu}}]{PhysRevB.91.214405}
\bibinfo{author}{\bibfnamefont{Y.}~\bibnamefont{Gao}},
  \bibinfo{author}{\bibfnamefont{S.~A.} \bibnamefont{Yang}}, \bibnamefont{and}
  \bibinfo{author}{\bibfnamefont{Q.}~\bibnamefont{Niu}},
  \bibinfo{journal}{Phys. Rev. B} \textbf{\bibinfo{volume}{91}},
  \bibinfo{pages}{214405} (\bibinfo{year}{2015}),
  \urlprefix\url{https://link.aps.org/doi/10.1103/PhysRevB.91.214405}.

\bibitem[{\citenamefont{Iskin}(2018)}]{PhysRevA.97.033625}
\bibinfo{author}{\bibfnamefont{M.}~\bibnamefont{Iskin}},
  \bibinfo{journal}{Phys. Rev. A} \textbf{\bibinfo{volume}{97}},
  \bibinfo{pages}{033625} (\bibinfo{year}{2018}),
  \urlprefix\url{https://link.aps.org/doi/10.1103/PhysRevA.97.033625}.

\bibitem[{\citenamefont{Kibble}(1979)}]{cmp/1103904831}
\bibinfo{author}{\bibfnamefont{T.~W.~B.} \bibnamefont{Kibble}},
  \bibinfo{journal}{Commun. Math. Phys.} \textbf{\bibinfo{volume}{65}},
  \bibinfo{pages}{189 } (\bibinfo{year}{1979}).

\bibitem[{\citenamefont{Xiao et~al.}(2010)\citenamefont{Xiao, Chang, and
  Niu}}]{RevModPhys.82.1959}
\bibinfo{author}{\bibfnamefont{D.}~\bibnamefont{Xiao}},
  \bibinfo{author}{\bibfnamefont{M.-C.} \bibnamefont{Chang}}, \bibnamefont{and}
  \bibinfo{author}{\bibfnamefont{Q.}~\bibnamefont{Niu}}, \bibinfo{journal}{Rev.
  Mod. Phys.} \textbf{\bibinfo{volume}{82}}, \bibinfo{pages}{1959}
  (\bibinfo{year}{2010}),
  \urlprefix\url{https://link.aps.org/doi/10.1103/RevModPhys.82.1959}.

\bibitem[{\citenamefont{Chen and Huang}(2021)}]{PhysRevResearch.3.L042018}
\bibinfo{author}{\bibfnamefont{W.}~\bibnamefont{Chen}} \bibnamefont{and}
  \bibinfo{author}{\bibfnamefont{W.}~\bibnamefont{Huang}},
  \bibinfo{journal}{Phys. Rev. Res.} \textbf{\bibinfo{volume}{3}},
  \bibinfo{pages}{L042018} (\bibinfo{year}{2021}),
  \urlprefix\url{https://link.aps.org/doi/10.1103/PhysRevResearch.3.L042018}.

\bibitem[{\citenamefont{Qi et~al.}(2006)\citenamefont{Qi, Wu, and
  Zhang}}]{PhysRevB.74.085308}
\bibinfo{author}{\bibfnamefont{X.-L.} \bibnamefont{Qi}},
  \bibinfo{author}{\bibfnamefont{Y.-S.} \bibnamefont{Wu}}, \bibnamefont{and}
  \bibinfo{author}{\bibfnamefont{S.-C.} \bibnamefont{Zhang}},
  \bibinfo{journal}{Phys. Rev. B} \textbf{\bibinfo{volume}{74}},
  \bibinfo{pages}{085308} (\bibinfo{year}{2006}),
  \urlprefix\url{https://link.aps.org/doi/10.1103/PhysRevB.74.085308}.

\bibitem[{\citenamefont{Bernevig et~al.}(2021)\citenamefont{Bernevig, Lian,
  Cowsik, Xie, Regnault, and Song}}]{PhysRevB.103.205415}
\bibinfo{author}{\bibfnamefont{B.~A.} \bibnamefont{Bernevig}},
  \bibinfo{author}{\bibfnamefont{B.}~\bibnamefont{Lian}},
  \bibinfo{author}{\bibfnamefont{A.}~\bibnamefont{Cowsik}},
  \bibinfo{author}{\bibfnamefont{F.}~\bibnamefont{Xie}},
  \bibinfo{author}{\bibfnamefont{N.}~\bibnamefont{Regnault}}, \bibnamefont{and}
  \bibinfo{author}{\bibfnamefont{Z.-D.} \bibnamefont{Song}},
  \bibinfo{journal}{Phys. Rev. B} \textbf{\bibinfo{volume}{103}},
  \bibinfo{pages}{205415} (\bibinfo{year}{2021}),
  \urlprefix\url{https://link.aps.org/doi/10.1103/PhysRevB.103.205415}.

\bibitem[{\citenamefont{Bhandari et~al.}(2020)\citenamefont{Bhandari, Alonso,
  Taddei, von Oppen, Fazio, and Arrachea}}]{PhysRevB.102.155407}
\bibinfo{author}{\bibfnamefont{B.}~\bibnamefont{Bhandari}},
  \bibinfo{author}{\bibfnamefont{P.~T.} \bibnamefont{Alonso}},
  \bibinfo{author}{\bibfnamefont{F.}~\bibnamefont{Taddei}},
  \bibinfo{author}{\bibfnamefont{F.}~\bibnamefont{von Oppen}},
  \bibinfo{author}{\bibfnamefont{R.}~\bibnamefont{Fazio}}, \bibnamefont{and}
  \bibinfo{author}{\bibfnamefont{L.}~\bibnamefont{Arrachea}},
  \bibinfo{journal}{Phys. Rev. B} \textbf{\bibinfo{volume}{102}},
  \bibinfo{pages}{155407} (\bibinfo{year}{2020}),
  \urlprefix\url{https://link.aps.org/doi/10.1103/PhysRevB.102.155407}.

\bibitem[{\citenamefont{Yu et~al.}(2019)\citenamefont{Yu, Yang, Gong, Cao, Lu,
  Liu, Zhang, Plenio, Jelezko, Ozawa et~al.}}]{10.1093/nsr/nwz193}
\bibinfo{author}{\bibfnamefont{M.}~\bibnamefont{Yu}},
  \bibinfo{author}{\bibfnamefont{P.}~\bibnamefont{Yang}},
  \bibinfo{author}{\bibfnamefont{M.}~\bibnamefont{Gong}},
  \bibinfo{author}{\bibfnamefont{Q.}~\bibnamefont{Cao}},
  \bibinfo{author}{\bibfnamefont{Q.}~\bibnamefont{Lu}},
  \bibinfo{author}{\bibfnamefont{H.}~\bibnamefont{Liu}},
  \bibinfo{author}{\bibfnamefont{S.}~\bibnamefont{Zhang}},
  \bibinfo{author}{\bibfnamefont{M.~B.} \bibnamefont{Plenio}},
  \bibinfo{author}{\bibfnamefont{F.}~\bibnamefont{Jelezko}},
  \bibinfo{author}{\bibfnamefont{T.}~\bibnamefont{Ozawa}},
  \bibnamefont{et~al.}, \bibinfo{journal}{National Science Review}
  \textbf{\bibinfo{volume}{7}}, \bibinfo{pages}{254} (\bibinfo{year}{2019}),
  ISSN \bibinfo{issn}{2095-5138},
  \eprint{https://academic.oup.com/nsr/article-pdf/7/2/254/38881669/nwz193\_supplemental\_file.pdf},
  \urlprefix\url{https://doi.org/10.1093/nsr/nwz193}.

\bibitem[{\citenamefont{Tan et~al.}(2019)\citenamefont{Tan, Zhang, Yang, Chu,
  Zhu, Li, Yang, Song, Han, Li et~al.}}]{PhysRevLett.122.210401}
\bibinfo{author}{\bibfnamefont{X.}~\bibnamefont{Tan}},
  \bibinfo{author}{\bibfnamefont{D.-W.} \bibnamefont{Zhang}},
  \bibinfo{author}{\bibfnamefont{Z.}~\bibnamefont{Yang}},
  \bibinfo{author}{\bibfnamefont{J.}~\bibnamefont{Chu}},
  \bibinfo{author}{\bibfnamefont{Y.-Q.} \bibnamefont{Zhu}},
  \bibinfo{author}{\bibfnamefont{D.}~\bibnamefont{Li}},
  \bibinfo{author}{\bibfnamefont{X.}~\bibnamefont{Yang}},
  \bibinfo{author}{\bibfnamefont{S.}~\bibnamefont{Song}},
  \bibinfo{author}{\bibfnamefont{Z.}~\bibnamefont{Han}},
  \bibinfo{author}{\bibfnamefont{Z.}~\bibnamefont{Li}}, \bibnamefont{et~al.},
  \bibinfo{journal}{Phys. Rev. Lett.} \textbf{\bibinfo{volume}{122}},
  \bibinfo{pages}{210401} (\bibinfo{year}{2019}),
  \urlprefix\url{https://link.aps.org/doi/10.1103/PhysRevLett.122.210401}.

\bibitem[{\citenamefont{Gianfrate et~al.}(2020)\citenamefont{Gianfrate, Bleu,
  Dominici, Ardizzone, De~Giorgi, Ballarini, Lerario, West, Pfeiffer,
  Solnyshkov et~al.}}]{Gianfrate2020}
\bibinfo{author}{\bibfnamefont{A.}~\bibnamefont{Gianfrate}},
  \bibinfo{author}{\bibfnamefont{O.}~\bibnamefont{Bleu}},
  \bibinfo{author}{\bibfnamefont{L.}~\bibnamefont{Dominici}},
  \bibinfo{author}{\bibfnamefont{V.}~\bibnamefont{Ardizzone}},
  \bibinfo{author}{\bibfnamefont{M.}~\bibnamefont{De~Giorgi}},
  \bibinfo{author}{\bibfnamefont{D.}~\bibnamefont{Ballarini}},
  \bibinfo{author}{\bibfnamefont{G.}~\bibnamefont{Lerario}},
  \bibinfo{author}{\bibfnamefont{K.~W.} \bibnamefont{West}},
  \bibinfo{author}{\bibfnamefont{L.~N.} \bibnamefont{Pfeiffer}},
  \bibinfo{author}{\bibfnamefont{D.~D.} \bibnamefont{Solnyshkov}},
  \bibnamefont{et~al.}, \bibinfo{journal}{Nature}
  \textbf{\bibinfo{volume}{578}}, \bibinfo{pages}{381} (\bibinfo{year}{2020}),
  ISSN \bibinfo{issn}{1476-4687},
  \urlprefix\url{https://doi.org/10.1038/s41586-020-1989-2}.

\bibitem[{\citenamefont{Yi et~al.}(2023)\citenamefont{Yi, Yu, Yuan, Jiao, Yang,
  Jiang, Zhang, Chen, and Pan}}]{Yi23}
\bibinfo{author}{\bibfnamefont{C.-R.} \bibnamefont{Yi}},
  \bibinfo{author}{\bibfnamefont{J.}~\bibnamefont{Yu}},
  \bibinfo{author}{\bibfnamefont{H.}~\bibnamefont{Yuan}},
  \bibinfo{author}{\bibfnamefont{R.-H.} \bibnamefont{Jiao}},
  \bibinfo{author}{\bibfnamefont{Y.-M.} \bibnamefont{Yang}},
  \bibinfo{author}{\bibfnamefont{X.}~\bibnamefont{Jiang}},
  \bibinfo{author}{\bibfnamefont{J.-Y.} \bibnamefont{Zhang}},
  \bibinfo{author}{\bibfnamefont{S.}~\bibnamefont{Chen}}, \bibnamefont{and}
  \bibinfo{author}{\bibfnamefont{J.-W.} \bibnamefont{Pan}},
  \emph{\bibinfo{title}{Extracting the quantum geometric tensor of an optical
  raman lattice by bloch state tomography}} (\bibinfo{year}{2023}),
  \bibinfo{note}{arXiv:2301.06090}.

\bibitem[{\citenamefont{Cuerda et~al.}(2023)\citenamefont{Cuerda, Taskinen,
  Kallman, Grabitz, and Torma}}]{Cuerda23}
\bibinfo{author}{\bibfnamefont{J.}~\bibnamefont{Cuerda}},
  \bibinfo{author}{\bibfnamefont{J.~M.} \bibnamefont{Taskinen}},
  \bibinfo{author}{\bibfnamefont{N.}~\bibnamefont{Kallman}},
  \bibinfo{author}{\bibfnamefont{L.}~\bibnamefont{Grabitz}}, \bibnamefont{and}
  \bibinfo{author}{\bibfnamefont{P.}~\bibnamefont{Torma}},
  \emph{\bibinfo{title}{Observation of quantum metric and non-hermitian berry
  curvature in a plasmonic lattice}} (\bibinfo{year}{2023}),
  \bibinfo{note}{arXiv:2305.13174}.

\bibitem[{\citenamefont{Ozawa and Goldman}(2018)}]{PhysRevB.97.201117}
\bibinfo{author}{\bibfnamefont{T.}~\bibnamefont{Ozawa}} \bibnamefont{and}
  \bibinfo{author}{\bibfnamefont{N.}~\bibnamefont{Goldman}},
  \bibinfo{journal}{Phys. Rev. B} \textbf{\bibinfo{volume}{97}},
  \bibinfo{pages}{201117} (\bibinfo{year}{2018}),
  \urlprefix\url{https://link.aps.org/doi/10.1103/PhysRevB.97.201117}.

\bibitem[{\citenamefont{Bleu et~al.}(2018{\natexlab{a}})\citenamefont{Bleu,
  Malpuech, Gao, and Solnyshkov}}]{PhysRevLett.121.020401}
\bibinfo{author}{\bibfnamefont{O.}~\bibnamefont{Bleu}},
  \bibinfo{author}{\bibfnamefont{G.}~\bibnamefont{Malpuech}},
  \bibinfo{author}{\bibfnamefont{Y.}~\bibnamefont{Gao}}, \bibnamefont{and}
  \bibinfo{author}{\bibfnamefont{D.~D.} \bibnamefont{Solnyshkov}},
  \bibinfo{journal}{Phys. Rev. Lett.} \textbf{\bibinfo{volume}{121}},
  \bibinfo{pages}{020401} (\bibinfo{year}{2018}{\natexlab{a}}),
  \urlprefix\url{https://link.aps.org/doi/10.1103/PhysRevLett.121.020401}.

\bibitem[{\citenamefont{Bleu et~al.}(2018{\natexlab{b}})\citenamefont{Bleu,
  Solnyshkov, and Malpuech}}]{PhysRevB.97.195422}
\bibinfo{author}{\bibfnamefont{O.}~\bibnamefont{Bleu}},
  \bibinfo{author}{\bibfnamefont{D.~D.} \bibnamefont{Solnyshkov}},
  \bibnamefont{and} \bibinfo{author}{\bibfnamefont{G.}~\bibnamefont{Malpuech}},
  \bibinfo{journal}{Phys. Rev. B} \textbf{\bibinfo{volume}{97}},
  \bibinfo{pages}{195422} (\bibinfo{year}{2018}{\natexlab{b}}),
  \urlprefix\url{https://link.aps.org/doi/10.1103/PhysRevB.97.195422}.

\bibitem[{\citenamefont{Klees et~al.}(2020)\citenamefont{Klees, Rastelli,
  Cuevas, and Belzig}}]{PhysRevLett.124.197002}
\bibinfo{author}{\bibfnamefont{R.~L.} \bibnamefont{Klees}},
  \bibinfo{author}{\bibfnamefont{G.}~\bibnamefont{Rastelli}},
  \bibinfo{author}{\bibfnamefont{J.~C.} \bibnamefont{Cuevas}},
  \bibnamefont{and} \bibinfo{author}{\bibfnamefont{W.}~\bibnamefont{Belzig}},
  \bibinfo{journal}{Phys. Rev. Lett.} \textbf{\bibinfo{volume}{124}},
  \bibinfo{pages}{197002} (\bibinfo{year}{2020}),
  \urlprefix\url{https://link.aps.org/doi/10.1103/PhysRevLett.124.197002}.

\bibitem[{\citenamefont{Hou et~al.}(2024)\citenamefont{Hou, Zhou, Wang, Guo,
  and Chien}}]{PhysRevB.110.035144}
\bibinfo{author}{\bibfnamefont{X.-Y.} \bibnamefont{Hou}},
  \bibinfo{author}{\bibfnamefont{Z.}~\bibnamefont{Zhou}},
  \bibinfo{author}{\bibfnamefont{X.}~\bibnamefont{Wang}},
  \bibinfo{author}{\bibfnamefont{H.}~\bibnamefont{Guo}}, \bibnamefont{and}
  \bibinfo{author}{\bibfnamefont{C.-C.} \bibnamefont{Chien}},
  \bibinfo{journal}{Phys. Rev. B} \textbf{\bibinfo{volume}{110}},
  \bibinfo{pages}{035144} (\bibinfo{year}{2024}),
  \urlprefix\url{https://link.aps.org/doi/10.1103/PhysRevB.110.035144}.

\bibitem[{\citenamefont{Zhou et~al.}(2024)\citenamefont{Zhou, Hou, Wang, Tang,
  Guo, and Chien}}]{PhysRevB.110.035404}
\bibinfo{author}{\bibfnamefont{Z.}~\bibnamefont{Zhou}},
  \bibinfo{author}{\bibfnamefont{X.-Y.} \bibnamefont{Hou}},
  \bibinfo{author}{\bibfnamefont{X.}~\bibnamefont{Wang}},
  \bibinfo{author}{\bibfnamefont{J.-C.} \bibnamefont{Tang}},
  \bibinfo{author}{\bibfnamefont{H.}~\bibnamefont{Guo}}, \bibnamefont{and}
  \bibinfo{author}{\bibfnamefont{C.-C.} \bibnamefont{Chien}},
  \bibinfo{journal}{Phys. Rev. B} \textbf{\bibinfo{volume}{110}},
  \bibinfo{pages}{035404} (\bibinfo{year}{2024}),
  \urlprefix\url{https://link.aps.org/doi/10.1103/PhysRevB.110.035404}.

\bibitem[{\citenamefont{Andersson and Heydari}(2014)}]{Andersson_2014}
\bibinfo{author}{\bibfnamefont{O.}~\bibnamefont{Andersson}} \bibnamefont{and}
  \bibinfo{author}{\bibfnamefont{H.}~\bibnamefont{Heydari}},
  \bibinfo{journal}{J. Phys. A: Math. Gen.} \textbf{\bibinfo{volume}{47}},
  \bibinfo{pages}{215301} (\bibinfo{year}{2014}),
  \urlprefix\url{https://dx.doi.org/10.1088/1751-8113/47/21/215301}.

\bibitem[{\citenamefont{Sj\"oqvist}(2020)}]{PhysRevResearch.2.013344}
\bibinfo{author}{\bibfnamefont{E.}~\bibnamefont{Sj\"oqvist}},
  \bibinfo{journal}{Phys. Rev. Res.} \textbf{\bibinfo{volume}{2}},
  \bibinfo{pages}{013344} (\bibinfo{year}{2020}),
  \urlprefix\url{https://link.aps.org/doi/10.1103/PhysRevResearch.2.013344}.

\bibitem[{\citenamefont{Bengtsson and Zyczkowski}(2006)}]{Bengtsson_book}
\bibinfo{author}{\bibfnamefont{I.}~\bibnamefont{Bengtsson}} \bibnamefont{and}
  \bibinfo{author}{\bibfnamefont{K.}~\bibnamefont{Zyczkowski}},
  \emph{\bibinfo{title}{Geometry of Quantum States: An Introduction to Quantum
  Entanglement}} (\bibinfo{publisher}{Cambridge University Press},
  \bibinfo{address}{Cambridge, UK}, \bibinfo{year}{2006}).

\bibitem[{\citenamefont{Roy}(2014)}]{PhysRevB.90.165139}
\bibinfo{author}{\bibfnamefont{R.}~\bibnamefont{Roy}}, \bibinfo{journal}{Phys.
  Rev. B} \textbf{\bibinfo{volume}{90}}, \bibinfo{pages}{165139}
  (\bibinfo{year}{2014}),
  \urlprefix\url{https://link.aps.org/doi/10.1103/PhysRevB.90.165139}.

\bibitem[{\citenamefont{Swanson}(1992)}]{Swanson_Book}
\bibinfo{author}{\bibfnamefont{M.~S.} \bibnamefont{Swanson}},
  \emph{\bibinfo{title}{Path integrals and quantum processes}}
  (\bibinfo{publisher}{Academic Press Inc.}, \bibinfo{address}{Boston, MA},
  \bibinfo{year}{1992}).

\bibitem[{\citenamefont{Scully and Zubairy}(1997)}]{Scully_Book}
\bibinfo{author}{\bibfnamefont{M.~O.} \bibnamefont{Scully}} \bibnamefont{and}
  \bibinfo{author}{\bibfnamefont{M.~S.} \bibnamefont{Zubairy}},
  \emph{\bibinfo{title}{Quantum Optics}} (\bibinfo{publisher}{Cambridge
  University Press}, \bibinfo{address}{Cambridge, UK}, \bibinfo{year}{1997}).

\bibitem[{\citenamefont{Kleinert}(2016)}]{Kleinert_Book}
\bibinfo{author}{\bibfnamefont{H.}~\bibnamefont{Kleinert}},
  \emph{\bibinfo{title}{Particles and quantum fields}}
  (\bibinfo{publisher}{World Scientific Publishing},
  \bibinfo{address}{Singapore}, \bibinfo{year}{2016}).

\bibitem[{\citenamefont{H$\ddot{\text{u}}$bner}(1992)}]{HubnerPLA92}
\bibinfo{author}{\bibfnamefont{M.}~\bibnamefont{H$\ddot{\text{u}}$bner}},
  \bibinfo{journal}{Phys. Lett. A} \textbf{\bibinfo{volume}{163}},
  \bibinfo{pages}{239} (\bibinfo{year}{1992}).

\end{thebibliography}

\end{document}